\newif\ifsingle

\ifsingle
\documentclass[11pt,draftclsnofoot, onecolumn]{IEEEtran}		
\else		
\documentclass[10pt,final, twocolumn]{IEEEtran}
\fi

\usepackage{times}
\usepackage{amsmath,dsfont}
\usepackage{amssymb,amsthm}
\usepackage{epsfig,verbatim}
\usepackage{subcaption}
\usepackage{setspace}
\usepackage{color}
\usepackage{cite}
\usepackage{epstopdf}
\usepackage{graphics}
\usepackage{accents}
\usepackage{acronym}
\usepackage[bookmarks,colorlinks]{hyperref}
\usepackage{booktabs}
\usepackage{mathtools}
\usepackage{float} 
\usepackage{algpseudocode}
\usepackage{enumitem}
\usepackage{bm}
\usepackage{colortbl}

\usepackage[ruled,linesnumbered,vlined]{algorithm2e}
\SetKwInput{KwData}{\textbf{Init}} 

\DeclareMathAlphabet\mathbfcal{OMS}{cmsy}{b}{n}

\newcommand{\sbrackets}[1]{\left[#1\right]}

\newcommand{\abs}[1]{\left\lvert#1\right\rvert}

\newcommand{\expecteds}[1]{\mathds{E}\sbrackets{#1}}

\newcommand{\gvec}[1]{\mathbf{#1}}

\newcommand{\ft}[1]{\mathcal{#1}}

\newcommand{\myScal}[1]{{\mathrm{#1}}}
\newcommand{\myVec}[1]{{\boldsymbol{#1}}}
\newcommand{\myMat}[1]{{\boldsymbol{#1}}}

\newcommand{\mySet}[1]{\mathcal{#1}}

\acrodef{soa}[SOA]{State-of-the-Art}
\acrodef{sp}[SP]{{signal processing}}
\acrodef{rt}[RT]{{real-time}}
\acrodef{rw}[RW]{real world}
\acrodef{mb}[MB]{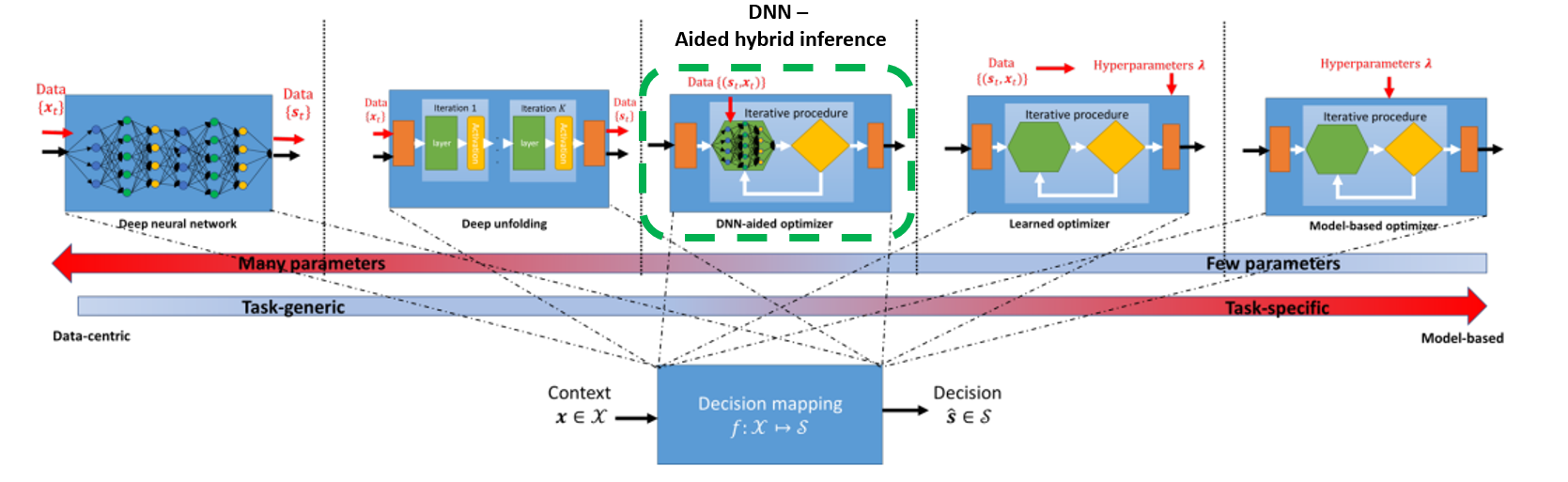}
\acrodef{dd}[DD]{data-driven}
\acrodef{e2e}[E2E]{end-to-end}
\acrodef{awgn}[AWGN]{additive white Gaussian noise}
\acrodef{lg}[LG]{linear Gaussian}
\acrodef{nl}[NL]{{non-linear}}

\acrodef{mse}[MSE]{mean-squared error}
\acrodef{rmse}[RMSE]{root mean squared error}
\acrodef{rmspe}[RMSPE]{root mean squared periodic error}
\acrodef{mmse}[MMSE]{{minimum mean-squared error}}
\acrodef{lmmse}[LMMSE]{{linear} MMSE}
\acrodef{mle}[MLE]{maximum likelihood estimation}
\acrodef{snr}[SNR]{signal-to-noise ratio}
\acrodef{ofdm}[OFDM]{orthogonal frequency-division multiplexing}

\acrodef{zzb}[ZZB]{Ziv-Zakai bound}

\acrodef{ml}[ML]{{machine learning}}
\acrodef{dl}[DL]{{deep learning}}
\acrodef{nn}[NN]{{neural network}}
\acrodef{dnn}[DNN]{{Deep neural network}}
\acrodef{rnn}[RNN]{{recurrent neural network}}
\acrodef{cnn}[CNN]{{convolutional neural network}}
\acrodef{dcnn}[DCNN]{{deconvolutional neural network}}
\acrodef{lstm}[LSTM]{{long short-term memory}}
\acrodef{gru}[GRU]{{gated recurrent unit}}
\acrodef{fc}[FC]{{Fully Connected}}
\acrodef{gt}[GT]{{Ground Truth}}
\acrodef{mlp}[MLP]{multi-layer perceptron}

\acrodef{sgd}[SGD]{stochastic gradient descent} 

\acrodef{doa}[DoA]{direction of arrival}
\acrodef{sv}[SV]{steering vector}
\acrodef{evd}[EVD]{eigenvalue decomposition}
\acrodef{sps}[SPS]{spatial smoothing}
\acrodef{fb}[FB]{forward-backward}

\acrodef{mprmse}[MPE]{minimal permutation root mean squared error}
\acrodef{music}[MUSIC]{Multiple Signal Classification}
\acrodef{mvdr}[MVDR]{minimum variance distortionless response}

\acrodef{rm}[Root-MUSIC]{Root-MUSIC}
\acrodef{esprit}[ESPRIT]{Estimation of Signal Parameters via Rotational Invariance Techniques}
\acrodef{drm}[DR-MUSIC]{Deep Root-MUSIC}
\acrodef{ssn}[SubspaceNet]{Subspace Net}

\acrodef{ula}[ULA]{uniform linear array}

\definecolor{Gray}{gray}{0.9}
\definecolor{LightCyan}{rgb}{0.88,1,1}

\newcommand{\secref}[1]{Section~\ref{#1}}

\IEEEoverridecommandlockouts
\title{SubspaceNet: Deep Learning-Aided Subspace Methods for DoA Estimation}

\author{
\IEEEauthorblockN{Dor H. Shmuel, Julian P. Merkofer, Guy Revach, Ruud J. G. van Sloun, and Nir Shlezinger
\thanks{
Parts of this work were presented in the 2023 IEEE International Conference on Acoustics Speech, and Signal Processing (ICASSP) as the paper \cite{shmuel2023deep}. 
 D. H. Shmuel and N. Shlezinger are with the School of ECE, Ben-Gurion University of the Negev, Beer Sheva, Israel (e-mail: shmueldo@post.bgu.ac.il; nirshl@bgu.ac.il).
 J. P. Merkofer and R. J. G. van Sloun are with the EE Dpt., Eindhoven University of Technology, The Netherlands (e-mail: \{j.p.merkofer; r.j.g.v.sloun\}@tue.nl).  
 G. Revach is with the D-ITET, ETH Zürich, Switzerland, (email: grevach@ethz.ch).
}}}

\begin{document}

\maketitle
\pagestyle{plain}
\thispagestyle{plain}
%
%
\begin{abstract} 
\Ac{doa} estimation is a fundamental task in array processing. A popular family of \ac{doa} estimation algorithms are subspace methods, which operate by dividing the measurements into distinct signal and noise subspaces. Subspace methods, such as \ac{music} and \acl{rm}, rely on several restrictive assumptions, including narrowband non-coherent sources and fully calibrated arrays, and their performance is considerably degraded when these do not hold. In this work we propose SubspaceNet; a \acl{dd} \ac{doa} estimator which learns how to divide the observations into distinguishable subspaces. 
This is achieved by utilizing a dedicated deep neural network to learn the empirical autocorrelation of the input, by training it as part of the \acl{rm} method, leveraging the inherent differentiability of this specific \ac{doa} estimator, while removing the need to provide a ground-truth decomposable autocorrelation matrix. Once trained, the resulting \acs{ssn} serves as a universal surrogate covariance estimator that can be applied in combination with any subspace-based \ac{doa} estimation method, allowing its successful application in challenging setups. \acs{ssn} is shown to  enable various \ac{doa} estimation algorithms to cope with coherent sources, wideband signals, low \acs{snr}, array mismatches, and limited snapshots, while preserving the interpretability and the suitability of classic subspace methods. 
%
%
\end{abstract}
%
%
\acresetall
\vspace{-0.2cm}
\section{Introduction}\label{sec:intro}
\vspace{-0.1cm}

%
\Ac{doa} estimation is a common array processing task, which deals with localizing emitting sources by determining the incidence angles of impinging waves~\cite{benesty2017fundamentals}.
A leading and well-trusted family of \ac{doa} estimators is based on {\em subspace methods}, including the celebrated \ac{music} algorithm~\cite{schmidt1986music}, \ac{rm}~\cite{Barabell1983ImprovingTR}, and \ac{esprit} \cite{ESPRIT}. When given sufficient signal snapshots, these techniques estimate multiple \acp{doa} with low complexity by separating the impinging wave into distinguishable signal and noise subspaces, achieving angular resolution that is not limited by the array geometry~\cite{benesty2017fundamentals}. 

The ability to separate an impinging wave into signal and noise subspaces, which is the core of subspace methods, relies on the orthogonality between the noise components and the array response. 
 {For this to hold, several key assumptions underlying subspace methods must be satisfied: 
$(i)$ the signals need to be narrowband; 
$(ii)$ the sources should be non-coherent; 
$(iii)$ the array has to be fully calibrated; 
$(iv)$ sufficient snapshots are needed such that one can faithfully to estimate the input covariance; 
and $(v)$ the statistical model underlying the estimation procedure has to be fully known. These requirements are often violated in practical settings, that frequently involve multiband signals, coherent sources (due to, e.g., multipath), miscalibrations in the array, limited snapshots, and mismatched models.}
Different pre-processing methods such as the \ac{sps} and \ac{fb}, have been proposed to tackle some of these limitations individually, e.g., \cite{wang1994spatial, 1518905}. These schemes average the power spectrum of the received signals over a number of adjacent sensor positions, thereby increasing the effective aperture, while resulting in a trade-off between resolution and variance reduction. Specifically,  larger averaging windows result in better variance reduction but poorer resolution, while smaller windows result in better resolution but poorer variance reduction. Hence, they tend to degrade the overall estimation performance and resolution, while constraining the number of recoverable sources.

Over the last decade, 
\acl{dl} has emerged as a leading tool for \acl{dd} inference. \acp{dnn}, which learn their mapping from data without relying on system modelling, have demonstrated unprecedented success in areas involving complex data such as computer vision~\cite{lecun2015deep}. 
%
%
%
%
%
%
Accordingly, \ac{dnn}-based approaches have recently been considered for \ac{doa} estimation~\cite{DNN_WITH_Antenna_ARRAY, cong2020robust, zhu2019deep,8170010,DOAEstimation_LowSNR,DOAEstimation_SparsePrior,lee2022deep, qin2023deep, lan2023novel, ResNET_singleSnapshot, DOAEstimation_HybridMIMOSystems, weisser2022unsupervised, Robust_DOA_estimation, elbir2020deepmusic,SingleSnapshot_CNN, lee2022ftmr, jiang2023toeplitz,  wu2022gridless, barthelme2021doa, ji2024transmusic, xu2024md,
merkofer2022deep,DA-MUSIC-2023}, as also surveyed in \cite{al2022review}. Specifically, in~\cite{DNN_WITH_Antenna_ARRAY, zhu2019deep, 8170010, DOAEstimation_LowSNR, ResNET_singleSnapshot, lan2023novel, lee2022deep, qin2023deep,  DOAEstimation_SparsePrior, cong2020robust} different \ac{dnn} architectures such as \acp{mlp} \cite{DNN_WITH_Antenna_ARRAY, cong2020robust}, \acp{cnn} \cite{zhu2019deep,8170010, DOAEstimation_LowSNR, DOAEstimation_SparsePrior, lee2022deep, qin2023deep},  attention models~\cite{lan2023novel}, and ResNet variants \cite{ResNET_singleSnapshot}, were trained to learn the mapping from the observations or their empirical covariance directly into \acp{doa}. 
While \acp{dnn} can be trained to estimate \acp{doa} without imposing specific requirements on the signal model, they typically require highly parameterized architectures trained with massive data sets, limiting their applicability on hardware-limited  devices where \ac{doa} estimation is often carried out. Furthermore, black-box \acp{dnn} lack the interpretability and flexibility of model-based algorithms, and often give rise to generalization issues.

An alternative approach uses \acp{dnn} alongside model-based \ac{doa} estimation, as a form of model-based deep learning~\cite{shlezinger2020model,shlezinger2022model}. For instance, the work \cite{DOAEstimation_HybridMIMOSystems} reduced the complexity of maximum likelihood \ac{doa} recovery by using a \ac{dnn} to limit the angular search space, thus still relying on accurate statistical modelling of the signals. 
 {
The works \cite{barthelme2021doa, wu2022gridless, jiang2023toeplitz} trained \ac{dnn} architectures to produce clean coavriance matrices that can be used for downstream subspace-based \ac{doa} estimation, while assuming access to a data set comprised of ideal covariances such that the \ac{dnn} can be trained separately. }
The work \cite{weisser2022unsupervised} used narrowband signal modelling to  train \ac{cnn}-based \ac{doa} estimators in an unsupervised manner. In the context of subspace methods,  \cite{Robust_DOA_estimation} trained an autoencoder to reconstruct a clean input from a corrupted  and then pass it to \ac{mlp} classifiers to reconstruct a \ac{music} spectrum,
while \cite{elbir2020deepmusic,SingleSnapshot_CNN, lee2022ftmr} used \acp{dnn} to directly produce a (discretized) spatial spectrum. However, these \ac{dnn}-aided estimators~\cite{Robust_DOA_estimation,elbir2020deepmusic,lee2022ftmr,SingleSnapshot_CNN} were not trained to provide \acp{doa}, and share the limitations of model-based subspace methods.
%
%

 {
Several recent works \cite{merkofer2022deep,DA-MUSIC-2023, ji2024transmusic, xu2024md} proposed architectures that jointly learns to compute the empirical covariance along with the processing of the resulting \ac{music} spectrum. The resulting estimators were capable of coping with, e.g., coherent and wideband sources.} However, this augmentation comes at the cost of compromising on the interpretability of subspace methods, as the \acp{doa} were recovered from the empirical covariance using an additional \ac{dnn}, learning a deep model where one cannot extract meaningful interpretation from its \ac{music} spectrum.  This motivates designing \ac{dnn}-aided \ac{doa} estimators that cope with the limitations of model-based subspace methods, e.g., coherent sources and few snapshots, while preserving their interpretability and suitability. 

Here, we propose {\em \acs{ssn}}, which leverages data to implement subspace-based \ac{doa} estimation. \acs{ssn} is designed to cope with coherent sources, broadband signals, few snapshots, and calibration mismatches. This is achieved by identifying that these limitations of model-based subspace methods are encapsulated in their reliance on the empirical covariance of the received signals. SubspaceNet thus preserves the operation of  subpsace methods while using a dedicated trainable autoencoder which learns from data how to map the empirical auto-correlation of the observed signal into a surrogate covariance that is universally useful for subspace-based \ac{doa} recovery, i.e., that is decomposable into signal and noise subspaces.

In order to train an autoencoder that produces subspace-compliant covariance estimates, we exploit the inherent differentiability of \acs{rm} to convert the algorithm into a trainable discriminative model~\cite{shlezinger2022discriminative}. By doing so, the trained \ac{dnn} computes meaningful subspace representations leading to  accurate \ac{doa} estimates, without having to provide ground-truth decomposable covariances.  
\acs{ssn} thus learns to compute meaningful subspace representations that, once trained, can be combined with different subspace \ac{doa} estimators, including \ac{music}, \acl{rm}, and \ac{esprit}.
%
We empirically show that SubspaceNet outperforms model-based subspace methods for both coherent and non-coherent sources, and that it successfully operates with broadband signals, calibration mismatches, low \ac{snr}, and few snapshots. We also demonstrate the SubspaceNet creates a clear and significant distinction between the noise and the signal subspaces, which facilitates its diagnosis and identifying of the number of sources. 

The rest of the paper is organized as follows: \secref{sec:Model} formulates the  setup and recalls subspace methods. \secref{sec:SubspaceNet} presents SubspaceNet, which is numerically  evaluated in \secref{sec:Numerical Evaluation}. Finally, \secref{ssec:conclusions} concludes the paper.

Throughout this paper, we use
boldface-uppercase for matrices, e.g., $\myVec{X}$, 
boldface-lowercase for vectors, e.g., $\myVec{x}$. 
We denote the $j$th entry of vector $\myVec{x}$ and the $(i,j)$th entry of matrix $\myMat{X}$ by $[\myVec{x}]_j$ and $[\myMat{X}]_{i,j}$, respectively.
We use $(\cdot)^{\mathsf{H}}$, $(\cdot)^T$, $\| \cdot\|$, and $ \expecteds{\cdot}$  for the hermitian transpose, transpose,  $\ell_2$ norm, and stochastic expectation, respectively.
\vspace{-0.2cm}
\section{System Model and Preliminaries}\label{sec:Model}
\vspace{-0.1cm}
In this section, we present the system model, beginning with received signal model in Subsection~\ref{ssec:Signal model}. We then briefly review subspace-based \ac{doa} estimation in Subsection~\ref{ssec:subspace_methods}, and formulate the considered problem in Subsection~\ref{ssec:Problem Formulation}.

\vspace{-0.2cm}
\subsection{\ac{doa} Estimation Signal Model}\label{ssec:Signal model}
\begin{figure}
    \centering
    \includegraphics[width=\columnwidth]{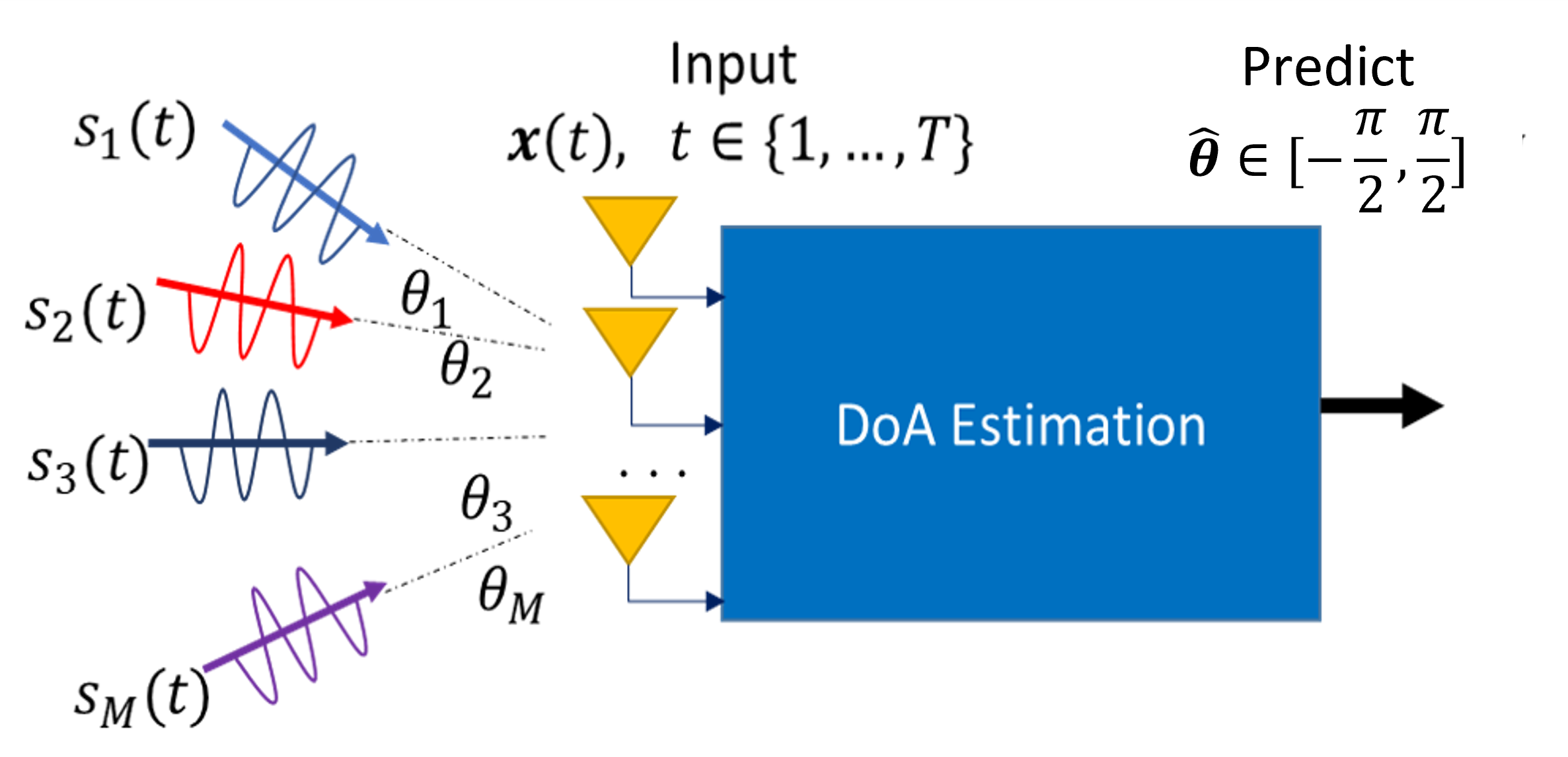}
    \caption{\ac{doa} estimation system illustration.}
    \label{fig:DoASysModel_1}
\end{figure}

\vspace{-0.1cm}
We consider a receiver equipped with a \ac{ula} composed of $N$ half-wavelength spaced antenna elements. At each discrete time instance $t$ (out of a horizon of $T$ snapshots, i.e., $t \in\{1,\hdots,T\}$), the receiver gathers multidimensional observations, denoted  $\myVec{x}(t) = [\myScal{x}_1(t),\hdots,\myScal{x}_N(t)] \in \mathbb{C}^{N}$. These signals originate from $M$ sources, represented as $\myVec{s}(t) = [\myScal{s}_1(t),\hdots,\myScal{s}_M(t)] \in \mathbb{C}^{M} $, and the \acp{doa} of these sources are denoted in vector form as $\myVec{\theta} = [\theta_1,  \hdots, \theta_M]$. The system is illustrated in Fig.~\ref{fig:DoASysModel_1}. {\em \ac{doa} estimation} is the recovery of $\myVec{\theta}$ from the observations obtained over $T$ snapshots.

When the transmitted signals are narrowband and in the far-field of the array, the  received signal over $T$ snapshots, denoted 
%
$\myMat{X} = [\myVec{x}(1), \ldots, \myVec{x}(T)] \in \mathbb{C}^{N\times T} $ is typically modelled as 
\begin{equation}\label{observation_formula}
    \myMat{X} = \myMat{A}(\myVec{\theta}) \myMat{S} + \myMat{V}.
\end{equation}
In \eqref{observation_formula}, $ \myMat{S} = [\myVec{s}(1),\ldots,\myVec{s}(T)] \in \mathbb{C}^{M\times T} $ is the sources signals matrix; $ \myMat{V} 
\in \mathbb{C}^{N\times T} $ is the noise matrix comprised of i.i.d. entries with variance $\sigma_V^2$; and 
$\myMat{A}(\myVec{\theta}) = [\myVec{a}(\theta_1),\ldots,\myVec{a}(\theta_M)] \in \mathbb{C} ^{N \times M}$ is the steering matrix, where
\begin{equation}
\label{eqn:SteeringVec}
    \myVec{a}(\theta) \triangleq [1, e^{-j\pi \sin(\theta)}, ... , e^{-j\pi(N-1)\sin(\theta)}].
\end{equation}

When the sources are zero-mean and stationary, the covariance of the observations is obtained from \eqref{observation_formula} as
\begin{equation}\label{covariance matrix equation}
    \myMat{R}_X = \expecteds{\myMat{X}\myMat{X}^\mathsf{H}} = \myMat{A}(\myVec{\theta}) \myMat{R}_S \myMat{A}^\mathsf{H}(\myVec{\theta}) + \sigma_V^2 \myMat{I}_{N},
\end{equation}
where $\myMat{R}_S$ denotes the covariance of $\myVec{S}$. 

%
%

\vspace{-0.2cm}
\subsection{Subspace-Based \ac{doa} Estimation}
\label{ssec:subspace_methods}
\vspace{-0.1cm}
Subspace methods are a family of \ac{doa} estimation algorithms that rely on the ability to decompose the observations covariance matrix in \eqref{covariance matrix equation} into orthogonal signal and noise subspaces. They are based on the following assumptions:

\begin{enumerate}[label={AS\arabic*}]
\item \label{itm:narrowband} The sources are narrowband, such that \eqref{observation_formula} holds.
\item \label{itm:coherent} The sources are non-coherent, namely, the covariance matrix  $\myMat{R}_S$ in \eqref{covariance matrix equation} is of full rank.
\item \label{itm:callib} The array is calibrated, such that the steering vectors in \eqref{eqn:SteeringVec} are known and accurately match the array.
\item \label{itm:Snapshots} The number of snapshots $T$ is sufficiently large, and the \ac{snr} is sufficiently high, such that one can reliably estimate $ \myMat{R}_X$ in \eqref{covariance matrix equation} empirically as
\begin{equation}
\label{eqn:empCov}
    \hat{\myMat{R}}_X = \frac{1}{T}\myMat{X}\myMat{X}^\mathsf{H}.
\end{equation}
\end{enumerate}


Under assumptions \ref{itm:narrowband}-\ref{itm:callib}, subspace methods decompose the covariance matrix $\myMat{R}_X$ in \eqref{covariance matrix equation}, which can be sufficiently estimated under \ref{itm:Snapshots}, into orthogonal subspaces. To see this, we write the \ac{evd} of $\myMat{R}_X$  as  
\begin{equation}\label{eqn:evd}
    \myMat{R}_X = \myMat{U}\myMat{\Lambda}\myMat{U}^\mathsf{H}, 
\end{equation}
where  $ \myMat{\Lambda}$ is the diagonal eigenvalues matrix while the unitary matrix $\myMat{U}= [\myVec{u}_1, \ldots,\myVec{u}_N]$ is their corresponding eigenvectors.
The resulting eigenvectors $\mathbf{U}$ span an observation space that can be divided into two orthogonal subspaces: the signal subspace $\mathbf{U}_S$ and the noise subspace $\mathbf{U}_N$~\cite{schmidt1986music}, i.e.,
\begin{equation}\label{orthogonal_subspaces}
    \myMat{U} = [ \myMat{U}_{\rm S} | \myMat{U}_{\rm N} ], \quad \text{where} \quad \myMat{U}_{\rm S} \bot \myMat{U}_{\rm N}.
\end{equation}
In \eqref{orthogonal_subspaces},  $ \myMat{U}_{\rm S}$ contains $M$ eigenvectors, each corresponding to a specific directional impinging source \cite{godara2004smart}, and is spanned by the columns of the steering matrix $\myMat{A}(\myVec{\theta})$; the $(N-M)\times N$ matrix $\myMat{U}_{\rm N}$, comprised of the eigenvectors corresponding to the $N-M$ least dominant eigenvalues of $\myMat{R}_X$ (which equal $\sigma_V^2$), represents the noise subspace.

The representation \eqref{orthogonal_subspaces} implies that   $\myMat{A}(\myVec{\theta}) \bot \myMat{U}_{\rm N}$, leading to 
\begin{equation}\label{basic_core_equation}
    \myVec{a}^\mathsf{H}({\theta}_i)\myMat{U}_{\rm N}\myMat{U}_{\rm N}^\mathsf{H}\myVec{a}
    ({\theta}_i) = 
    \left\| \myMat{U}_{\rm N}^\mathsf{H}\myVec{a}({\theta}_i) \right\|^2 =  0,
\end{equation}
for all $ i\in 1,\ldots,M$.
Equation \eqref{basic_core_equation} forms the basis of subspace methods, which use it to identify the \acp{doa}, as reviewed next.

\subsubsection{\ac{music}} Arguably the most common subspace-based \ac{doa} estimator is  \ac{music} \cite{schmidt1986music}. Here,  \eqref{basic_core_equation} is used to construct a spectrum representation by taking the inverse projection of the steering vectors on the noise subspace. In particular, the noise subspace matrix and the number of sources are estimated from the \ac{evd} of the empirical covariance \eqref{eqn:empCov} as $\hat{M}$ and $\hat{\myMat{U}}_{N}$, respectively.  The resulting \ac{music} spectrum is given by 
\begin{equation}\label{MUSIC_spectrum}
    P_{\rm MUSIC}(\theta) = \frac{1}{\big\| {\hat{\myMat{U}}}_{N}^\mathsf{H}\myVec{a}(\theta)\big\|^2}.
\end{equation}
The \acp{doa} are then recovered by identifying the peaks of the constructed spectrum $ P_{\rm MUSIC}(\theta)$, i.e., the \ac{music} spectrum provides an interpretable visual representation of the \acp{doa}.

\subsubsection{\acs{rm}} An alternative approach to recover the \acp{doa} from the estimated   $\hat{\myMat{U}}_{\rm N}$ seeks the roots of \eqref{basic_core_equation}.
In particular, \acs{rm} formulates the Hermitian matrix $\myMat{F}=\hat{\myMat{U}}_{\rm N}\hat{\myMat{U}}_{\rm N}^\mathsf{H}$ using its diagonal sum coefficients ${f_n}$, defined as
\begin{equation}\label{eq:diagonal_coeff}
    f_n = \sum_{i=0}^{N-1-n}[\myMat{F}]_{i,n+i}, \> \> \> n \geq 0,
\end{equation}
where for $n < 0, f_n = f_{\abs{n}}^\ast$. Substituting \eqref{eq:diagonal_coeff} into \eqref{basic_core_equation}, the left hand side of \eqref{basic_core_equation} can be expressed as a polynomial equation in a complex-valued argument $z$ of order $2N-2$, given by
\begin{align} 
    D(z)&= \sum_{i=0}^{N-1}\sum_{j=0}^{N-1}[\myVec{a}(\theta)]_i^\ast[\myMat{F}]_{ij}[\myVec{a}(\theta)]_j  
    \notag  \\
    &=\sum_{i=0}^{N-1}\sum_{j=0}^{N-1}{[\myMat{F}]_{ij}z^{i-j}}
    =
    \sum_{n=-(N-1)}^{N-1}{f_n z^n},
    \label{eqn:polynomial}
\end{align}
where $z=e^{-j\pi \sin(\theta)}$. \acs{rm} identifies the \acp{doa} from the roots of the polynomial \eqref{eqn:polynomial}. The roots map is viewed as the \acs{rm} spectrum. Since \eqref{eqn:polynomial} has $2N-2 > M$ roots (divided into symmetric pairs), while the roots corresponding to \acp{doa} should have unit magnitude, the $\hat{M}$ pairs of roots which are the closest to the unit circle are matched as the $\hat{M}$ sources \acp{doa} \cite{Barabell1983ImprovingTR}. Similarly to \ac{music}, the \acs{rm} spectrum
provides a meaningful visualization of the \acp{doa}. 
\subsubsection{\ac{esprit}} The popular \ac{esprit} subspace method separates the array into two overlapping sub-arrays, where it exploits the rotational invariance of the covariance matrix eigen-structure to simplify \ac{doa} estimation \cite{ESPRIT}. The steering matrices of the subarrays are related by $\myMat{A}_2(\theta) = \myMat{A}_1(\theta)\myMat{E}$, where $\myMat{E}$ is a diagonal matrix with entries $[\myMat{E}]_{m,m} = e^{-j\pi\sin(\theta_m)}$.
Since $\myMat{A}(\theta)$ spans the signal subspace, the relation between the signal subspaces of the sub-arrays, which we denote by $\myMat{U}_{S_1}$ and $\myMat{U}_{S_2}$  for $\myMat{A}_1(\theta)$ and $\myMat{A}_2(\theta)$, respectively, can be expressed as
\begin{equation}\label{ESPRIT equation}
    \myMat{U}_{S_2} = \myMat{U}_{S_1}\myMat{H},
\end{equation}
for some matrix $\myMat{H}$.
Since the eigenvalues of $\myMat{H}$ are equal to the diagonal elements of $\myMat{E}$, by solving \eqref{ESPRIT equation}, i.e., extracting $\myMat{H}$, one can recover the  \acp{doa}  from these eigenvalues. Consequently, \ac{esprit} estimates the \ac{doa} vector as $\hat{\myVec{\theta}} = \sin^{-1}(\angle{{\rm eig}\{\myMat{H}\}} / {\pi})$, where ${\rm eig}\{\myMat{H}\}$ is the vector representation of the eigenvalues of $\myMat{H}$, and $\angle(\cdot)$ is the angle operator.  



%
\vspace{-0.2cm}
\subsection{Problem Formulation}
\label{ssec:Problem Formulation}
\vspace{-0.1cm}
Subspace methods reliably estimate  \acp{doa} with an angular resolution that is not limited by the physical array, and they typically provide an interpretable visualization of  the \acp{doa} along with uncertainty measures. However, their reliance on Assumptions \ref{itm:narrowband}-\ref{itm:Snapshots} can be a limiting factor in various practical scenarios, which often involve broadband signals, coherent sources, miscalibrations, limited snapshots, and low \ac{snr}. Our goal is to design a method which enables subspace methods to operate also when \ref{itm:narrowband}-\ref{itm:Snapshots} do not hold.

 In particular, we consider the  recovery of $\myVec{\theta}$ from the observations matrix $\myMat{X}$. While the number of sources $M$ is not assumed to be a-priori known, we limit our attention to settings where $M$ is smaller than the number of antenna elements $N$. Similarly, the number of snapshots $T$ is not known during system design, though it is lower bounded by some $T_{\min}$ (which can be as small as  a single snapshot).
 
To alleviate reliance on \ref{itm:narrowband}-\ref{itm:Snapshots}, we consider a data-aided setting, where one has access to a data set comprised of $J$ pairs of observations and their corresponding \acp{doa}, denoted
\begin{equation}
\label{eqn:Dataset}
  \mySet{D} = \left\{\left(\myMat{X}^{(j)}, \myVec{\theta}^{(j)}\right)\right\}_{j = 1}^J.   
\end{equation}
This data can be obtained from real-world sources or from synthetic data that closely resembles the signals that are expected to be detected. 
Different realizations can contain different numbers of sources $M$ and different number of snapshots $T$, i.e., the dimensionality of $\myVec{\theta}^{(j)}$ and the number of columns of $\myMat{X}^{(j)}$ can vary with the sample index $j$.

%
\vspace{-0.2cm}
\section{SubspaceNet}
\label{sec:SubspaceNet}
\vspace{-0.1cm}
In this section we present the proposed \acs{ssn}. 
We begin by discussing the high-level rationale of the algorithm and its architecture in Subsections~\ref{ssec:HLdesign}-\ref{ssec:Architecture}, respectively. Then, we present how \acs{ssn} is trained and applied during inference in Subsections~\ref{ssec:Training}-\ref{ssec:Inference}, respectively, and provide a discussion in Subsection~\ref{ssec:discussion}. 

\begin{figure*}
\begin{center}
\begin{subfigure}[pt]{\linewidth}
\centering
\includegraphics[width=0.85\linewidth]{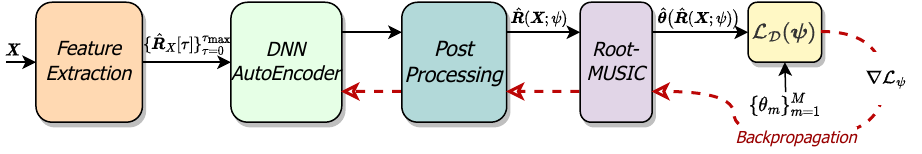}
\caption{\acs{ssn} training scheme}
\label{fig:ssn training scheme}
\end{subfigure}
\begin{subfigure}[pt]{\linewidth}
\centering
\includegraphics[width=0.8\linewidth]{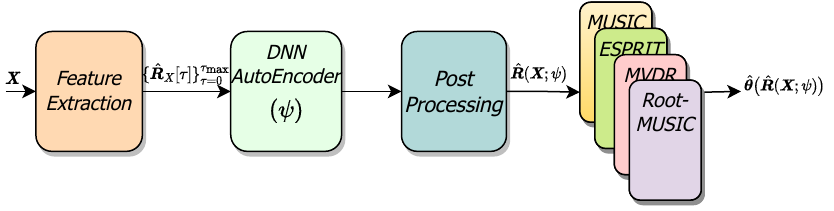}
\caption{\acs{ssn} inference scheme}
\label{fig:ssn inference scheme}
\end{subfigure}
\caption{High level illustration of \acs{ssn}}
\label{fig:ssn high-level scheme}
\vspace{-0.2cm}
\end{center}
\end{figure*}

%
\vspace{-0.2cm}
\subsection{High-Level Rationale}
\label{ssec:HLdesign}
\vspace{-0.1cm}
 The reason that Assumptions \ref{itm:narrowband}-\ref{itm:Snapshots} are critical for subspace methods follows from their role in the derivation of the orthogonality equality in \eqref{basic_core_equation}. When these assumptions do not hold, one cannot reliably obtain an estimate of the input covariance that admits a decomposition into orthogonal subspaces corresponding to the signals and the noise.
Based on the understanding that the sensitivity of subspace methods is encapsulated in the empirical covariance computation, we propose \acs{ssn}. \acs{ssn} augments subspace methods with deep learning as a form of model-based deep learning~\cite{shlezinger2020model,shlezinger2022model}, tackling the aforementioned challenges by providing a {\em surrogate} covariance matrix that can be divided into orthogonal signal and noise subspaces. To that aim, we employ a dedicated \ac{dnn} that learns from data to produce the desired covariance. The architecture of the \ac{dnn} is detailed in Subsection~\ref{ssec:Architecture}. This surrogate covariance can then be used by different subspace methods, preserving their operation and interpretability, as discussed in Subsection~\ref{ssec:Inference}.

Since there is no 'ground truth' surrogate covariance, the \ac{dnn} cannot be trained by comparing its output to a reference label, as commonly done in supervised learning. Instead, we evaluate the covariance produced by the \ac{dnn} based on its usefulness for subspace-based \ac{doa} estimation. Consequently, the training loss is computed by comparing the  \acp{doa} available in \eqref{eqn:Dataset} to the recovered \acp{doa} at the output of the subspace \ac{doa} estimator employing the \ac{dnn}, thus converting the \ac{doa} estimation algorithm into a  discriminative model~\cite{shlezinger2022discriminative} that is trainable end-to-end. Since \ac{dnn} training is based on first-order methods, the \ac{doa} estimator that processes the learned covariance during training has to be differentiable, i.e., one should be able to compute the gradient of the estimated \acp{doa} with respect to the estimated covariance. Unlike \ac{music}, whose mapping is based on  non-differentiable peak-finding, \ac{rm} is inherently differentiable, and is thus employed for training \acs{ssn}, as discussed in Subsection~\ref{ssec:Training}. 
%
The overall high-level design is presented in Fig.~\ref{fig:ssn high-level scheme}.

%
\vspace{-0.2cm}
\subsection{Architecture}
\label{ssec:Architecture}
\vspace{-0.1cm}

The trainable architecture employed by \acs{ssn} is designed to be able to map the input signal $\myMat{X}$ into an estimate of the covariance, denoted $\hat{\myMat{R}}$, regardless of the number of snapshots $T$. This is done in three stages: extracting features from $\myMat{X}$ that are informative for the surrogate covariance estimation task; processing these features using a \ac{dnn}-based autoencoder; and post-processing the \ac{dnn} outputs to obtain an estimated covariance. We next elaborate on these stages. 

\subsubsection{Feature Extraction}
Our design of the features used for computing the surrogate covariance draws inspiration from focusing techniques~\cite{yoon2006tops}. These approaches aim to convert broadband \ac{doa} recovery setups into surrogate narrowband ones by estimating the covariance as a linear combination of the input multivariate power spectral density. Since the power spectral density of a stationary signal is  obtained as the Fourier transform of its auto-correlation function \cite{stoica2005spectral}, we use the empirical auto-correlation of $\myVec{x}(t)$ as our input features. 

In particular, we fix a maximal number of lags $\tau_{\max} >0$, which is not larger than the minimal number of snapshots $T_{\min}$. Then,  for each $\tau \in \{0.\ldots, \tau_{\max}\}$, we compute the   empirical auto-correlation of $\myMat{X}$ as 
\begin{equation}
    \label{eqn:AutoCorr}
    \hat{\myMat{R}}_X[\tau] = \frac{1}{T-\tau}\sum_{t=1}^{T-\tau}\myVec{x}(t)\myVec{x}^\mathsf{H}(t+\tau).
\end{equation}
The extracted features, processed by the \ac{dnn} detailed next, are thus an $N\times N \times (\tau_{\max}+1)$ tensor, i.e., its dimensionality does not depend on $T$. Consequently, this both provides the \ac{dnn} with features that are considered to be informative for obtaining a useful surrogate covariance matrix, while allowing the \ac{dnn} to be invariant of the number of snapshots $T$.

\begin{figure}
\centering
\includegraphics[width=0.8\columnwidth]{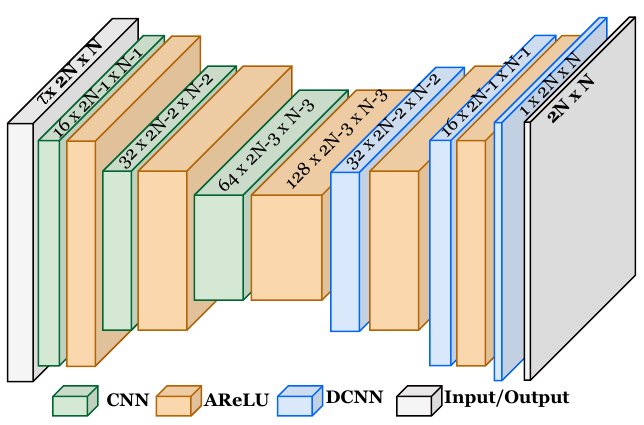}
\caption{\acs{ssn} autoencoder architecture}
\label{fig:ssn AE scheme}
\end{figure}

\subsubsection{\ac{dnn} Autoencoder}
\label{sssec:Autoenc}
We utilize a \ac{dnn} architecture inspired by denoising autoencoders to process   $\{\hat{\myMat{R}}_X[\tau]\}_{\tau=0}^{\tau_{\max}}$, as illustrated in Fig.~\ref{fig:ssn AE scheme}.  {Autoencoders are deep learning architectures that are  used for selecting relevant features from noisy data and mapping them into high-dimensional representations \cite[Ch. 14]{Goodfellow_Deep_Learning}.}
The encoder part of the \ac{dnn} consists of three \ac{cnn} layers  gradually increasing filter size. Its goal is to map the input into a lower dimensional latent representation that preserves only the meaningful information for the task, thus filtering out unnecessary information and noise. The decoder, which is comprised of a decreasing filter size \ac{dcnn} layers, aims to recover successive information details from the latent features as it progresses through its layers, producing a $2N\times N$ matrix which is post-processed into the surrogate covariance.

We use \ac{cnn} layers with $2\times 2$ kernels. This small kernel size allows to capture local phase correlations between neighboring sensors over the auto-correlation input, and potentially improve robustness to noise. The latter follows since the \ac{snr} across the spectrum is not constant, and thus the filters can detect local phase structures from the high \ac{snr} spectral regions to compensate for the lack of information from the low \ac{snr} regions. Unlike the conventional usage of \ac{cnn} autoencoders in, e.g., computer vision, we do not employ pooling layers to avoid discarding useful spatial information. 
Instead of using common rectified linear unit  activations ${\rm ReLU}(z) = \max(0,z)$, we utilize  anti-rectifier activations.
This ensures that both positive and negative values are propagated, making it more suitable for sensory data containing both positive and negative values~\cite{ABLE}. The anti-rectifier activation is expressed as the concatenation of both the positive and the negative part of the input. 
Specifically, for an $N_z \times 1$ input vector $\myVec{z}=[z_1, \ldots z_{N_z}]^T$, the activation outputs a $2N_z \times 1$ vector   given by:
\begin{align}
    {\rm AReLU}(\myVec{z}) 
    &= {\rm AReLU}\left([z_1, \ldots z_{N_z}]^T\right) \notag \\
    &=
    \Big[{\rm ReLU}(z_1), \ldots  {\rm ReLU}(z_{N_z}), \notag \\
    &\qquad 
    {\rm ReLU}(-z_1), \ldots  {\rm ReLU}(-z_{N_z})\Big]^T.
    \label{AReLU}
\end{align}
\color{black}

\subsubsection{Post-Processing}
The output of the autoencoder is a  $2N\times N$ real matrix. To convert it into a surrogate covariance matrix, which is an $N\times N$ complex Hermitian positive definite matrix, we first reshape the output into a complex matrix $\myMat{K}\in\mathbb{C}^{N\times N}$, by treating the first $N$ rows as the real part and the remaining $N$ rows as the imaginary part. To formulate the surrogate covariance, $\myMat{K}$ is converted into a Hermitian positive definite covariance estimate via 
\begin{equation}
\label{eqn:SurrogateCov}
    \hat{\myMat{R}} =  \myMat{K}\myMat{K}^\mathsf{H}+ \epsilon \myMat{I}_N,
\end{equation}
where $\epsilon>0$ is a fixed hyperparameter of the architecture. 

The trainable parameters of the architecture are the weights of the autoencoder, denoted $\myVec{\psi}$. We thus write the overall mapping as $\hat{\myMat{R}}(\myMat{X};\myVec{\psi})$, encompassing the feature extraction, \ac{dnn} processing, and output reshaping stages. 

\vspace{0.2cm}
\subsection{Training}
\label{ssec:Training}
The training procedure uses the dataset $\mySet{D}$ \eqref{eqn:Dataset} to tune the parameters $\myVec{\psi}$ such that the surrogate covariance $\hat{\myMat{R}}(\myMat{X};\myVec{\psi})$ is useful for subspace-based \ac{doa} recovery, i.e., it can be decomposed into orthogonal subspaces encapsulating the signals and the noise. This is achieved by encouraging $\hat{\myMat{R}}(\myMat{X};\myVec{\psi})$ to be a covariance matrix using which the \acp{doa} recovered by \ac{rm}, denoted $\hat{\myVec{\theta}}\big(\hat{\myMat{R}}(\myMat{X};\myVec{\psi})\big)$, is an accurate estimate of $\myVec{\theta}$. To describe how this is carried out, we first discuss the loss measure, after which we explain the training procedure.


\subsubsection{Loss Measure}
The loss function evaluates the quality of \ac{doa} estimation by comparing the \acp{doa} recovered by \ac{rm} combined with \acs{ssn}.  While in principle one can utilize standard loss measures for regression tasks, e.g., mean-squared error, when evaluating \ac{doa} recovery, one has to account for the following properties:
\begin{enumerate}[label={P\arabic*}]
    \item \label{itm:periodicity} \acp{doa} are periodic quantities, i.e., estimating $\hat{\myVec{\theta}} + 2\pi$ is the same as  $\hat{\myVec{\theta}}$. Thus the difference between an angle and its estimate does not necessarily correspond to actual error. 
    \item \label{itm:permute} \ac{doa} estimation is invariant to the order of the predicted \acp{doa}, e.g., estimating $\hat{\myVec{\theta}} = [\theta_1, \theta_2]^T$  is equivalent to outputting $\hat{\myVec{\theta}} = [\theta_2, \theta_1]^T$.
    \item \label{itm:numdoas} The number of signals $\hat{M}$ is also estimated, and may thus differ from the true $M$, i.e., the cardinality of the output may not be equal to that of the ground-truth \acp{doa}.
\end{enumerate}

To account for \ref{itm:periodicity}, we employ the \ac{rmspe} loss \cite{Tirza_6062425}, which evaluates the element-wise error over the periodic range. We extend this measure to be permutation invariant by \ref{itm:permute}, by modifying the \ac{rmspe} loss to compare all possible combinations of predicted \acp{doa} with the ground truth labels. To cope with \ref{itm:numdoas}, we treat cases where $\hat{M} < M$ by setting the last $M-\hat{M}$ \acp{doa} recovered om $\hat{\myVec{\theta}}$ to be zero when evaluating the loss. 

The resulting loss function is used to evaluate \acs{ssn} when applied with \ac{rm} to input $\myMat{X}$ originating from \acp{doa} $\myVec{\theta}$ while accounting for \ref{itm:periodicity}-\ref{itm:numdoas}. It is  given by
\begin{align*}
    l(\myMat{X},\myVec{\theta};\myVec{\psi}) = \min_{\myMat{P} \in \mySet{P}} 
   \bigg( \frac{1}{M}\Big\| {\rm mod}_{\pi}\Big(\myVec{\theta} 
   &- \myMat{P}\hat{\myVec{\theta}}\big(\hat{\myMat{R}}(\myMat{X};\myVec{\psi})\big)\Big)\Big\|^2\bigg)^{\frac{1}{2}},
\end{align*}
where ${\rm mod}_{\pi}$ denotes the modulo operation with respect to the angle range of interest, i.e., $[-\pi/2, \pi/2]$, and $\mySet{P}$ is the set of all $M \times M$ permutation matrices, i.e., binary matrices where each row and column contains a single non-zero entry. 
The loss evaluated over a dataset $\mySet{D}$ is computed as
\begin{align}
    \mySet{L}_{\mySet{D}} (\myVec{\psi})= \frac{1}{|\mySet{D}|}\sum_{(\myMat{X}^{(j)},\myVec{\theta}^{(j)}) \in \mySet{D}} l(\myMat{X}^{(j)},\myVec{\theta}^{(j)};\myVec{\psi}),
    \label{eqn:RMSPE}
\end{align}




\subsubsection{Training Procedure}

The selection of \acs{rm} as our model-based estimator for training \acs{ssn}, is attributed to the differentiability of its mapping. Specifically, the \ac{doa} estimates are obtained from the roots of the polynomial equation \eqref{eqn:polynomial}, which can be expressed as an implicit polynomial function. Implicit differentiation makes it possible to compute the derivative of the polynomial roots, that are used for the estimated \acp{doa} in \ac{rm}, with respect to its coefficients $\{f_n\}$~\cite{bolte2021nonsmooth}. These coefficients $\{f_n\}$ are obtained from the \ac{evd} of the estimated covariance matrix, and are thus differentiable with respect to it, see, e.g.,~\cite{solomon2019deep}. 

The above property allows end-to-end optimization of the \ac{dnn} parameters of \acs{ssn} using first-order methods, e.g., \ac{sgd} and its variants. In particularl, automatic differentiation engines, e.g., Autograd~\cite{paszke2017automatic}, can compute the gradient of the loss \eqref{eqn:RMSPE} with respect to the trainable parameters $\myVec{\psi}$ by backpropagating through the \acs{rm} block in Fig.~\ref{fig:ssn training scheme}. 
%
%
The training procedure of \acs{ssn} using  \ac{sgd} is detailed as Algorithm~\ref{alg:Training algorithm}.
\begin{algorithm}
\caption{\acs{ssn} Training using \ac{sgd}}
\label{alg:Training algorithm} 
\SetKwInOut{Initialization}{Init}
\Initialization{Lags $\tau_{\max}$, hyperapameter $\epsilon$,                                               learning rate $\mu $, number of batches $B$, 
epochs number $e_{\max}$.}
\SetKwInOut{Input}{Input}
\Input{Training set $\mySet{D}$}  
{ 
    %
    %
    Initialize \ac{cnn} weights $\myVec{\psi}$\;
    \For{${\rm epoch} = 0, 1, \ldots, e_{\max}$}{%
        Randomly divide  $\mySet{D}$ into $B$ batches $\{\mySet{D}_b\}_{b=1}^B$\;
        \For{$b =1,2,\ldots B$}{
            \For{$(\myMat{X}_b^{(j)},\myVec{\theta}_b^{(j)})\in \mySet{D}_b$}
            {
                Compute $\hat{\myMat{R}}(\myMat{X}_b^{(j)};\myVec{\psi})$ with \acs{ssn}\;
                Use \ac{rm} to get $\hat{\myVec{\theta}}\big(\hat{\myMat{R}}(\myMat{X}_b^{(j)};\myVec{\psi})\big)$\;
            }
            Compute $\mySet{L}_{\mySet{D}_b}(\myVec{\psi})$ using \eqref{eqn:RMSPE}\;
            Update weights via $\myVec{\psi} \leftarrow \myVec{\psi}-\mu \nabla_{\myVec{\psi}}\mySet{L}_{\mySet{D}_b}(\myVec{\psi})$\;
            }

            


            
    } 
    \KwRet{$\myVec{\psi}$}
}
\end{algorithm}

\vspace{-0.2cm}
\subsection{Inference}
\label{ssec:Inference}
\vspace{-0.1cm}
We employ \ac{rm} during training due to its differentiability. However, once trained, \acs{ssn}  can be seamlessly incorporated into any classical subspace methods such as \ac{music} and \ac{esprit}, producing a covariance matrix which facilitates \ac{doa} estimation, as presented in Fig. \ref{fig:ssn inference scheme}.  The inference algorithm is summarized in Algorithm \ref{alg:Inference algorithm}.
%
This plug-and-play feature is a significant advantage, as it can notably enhance the performance of classical methods and enable them to operate reliably in harsh settings where \ref{itm:narrowband}-\ref{itm:Snapshots} do not hold, i.e., where the original \ac{mb} algorithms typically fail. This is thanks to the fact that the learned surrogate covariance matrix enables the algorithms to operate as if \ref{itm:narrowband}-\ref{itm:Snapshots}  are satisfied, thus they can precisely distinguish between the noise and signal subspaces, as we demonstrate in Section~\ref{sec:Numerical Evaluation}.

While the presentation of \acs{ssn} so far considers subspace-based \ac{doa} estimation, the fact that it learns a covariance that facilitates \ac{doa} recovery can also be utilized by other forms of covariance-based methods. For instance, in our numerical study in Section~\ref{sec:Numerical Evaluation} we show that the covariance computed by \acs{ssn} can  be utilized by \ac{mvdr} \cite{capon1969high}, which processes the covariance representation without requiring its decomposition into signal and noise subpace. We demonstrate that combining \ac{mvdr} with \acs{ssn} enables coping with settings where the beamformer typically struggles, e.g., coherent sources.

  \begin{algorithm}
    \caption{\acs{ssn} Inference}
    \label{alg:Inference algorithm} 
    \SetKwInOut{Initialization}{Init}
    \Initialization{Trained \acs{ssn} parameters $\myVec{\psi}$,\\ model-based subspace \ac{doa} estimator}
    \SetKwInOut{Input}{Input} 
    \Input{Observations $\myMat{X}$;}
    {    
      Apply  \acs{ssn} to $\myMat{X}$ to obtain $\hat{\myMat{R}}(\myMat{X};\myVec{\psi})$\;
      Apply subspace method to estimate $\hat{\myVec{\theta}}$ from $\hat{\myMat{R}}(\myMat{X};\myVec{\psi})$\;
    }\KwRet{$\hat{\myVec{\theta}}$}
\end{algorithm}


 \vspace{-0.2cm}
\subsection{Discussion}\label{ssec:discussion}
\vspace{-0.1cm}
The proposed \acs{ssn} leverages data by using deep learning tools to enable subspace methods to operate reliably in the presence of coherent sources, broadband signals, limited snapshots, miscalibrated arrays, and low \ac{snr}. This is achieved by identifying that the difficulty in dealing with these challenges is associated with the computation of the covariance matrix. Thus, we propose a dedicated \ac{dnn} architecture which learns to provide a surrogate covariance, by training its output to be useful for \ac{rm}-based \ac{doa} recovery.  {The resulting \acs{ssn} thus combines abstract \acp{dnn} that enable downstream subspace-based \ac{doa} recovery to be carried out accurately, while preserving the interpretable operation of these model-based methods, e.g., the ability to extract a meaningful spectrum representation, as consistently demonstrated in Section~\ref{sec:Numerical Evaluation}.}

As opposed to \cite{merkofer2022deep}, which focused on \ac{music} and had to modify its operation for training, the inherit differentiability of \acs{rm} allows to train the architecture end-to-end without affecting how the computed covariance is processed. 
 {Once trained, \acs{ssn} allows subspace methods to achieve accurate \ac{doa} estimation in harsh conditions, while also producing a meaningful spectrum, as demonstrated in Section~\ref{sec:Numerical Evaluation}.
This principled \ac{dnn} augmentation notably enhances subspace-based \ac{doa} recovery. Still, it induces tradeoffs in terms of complexity and adaptivity in some settings. For once, while \acs{ssn} is notably less complex compared to alternative data-driven methods~\cite{DA-MUSIC-2023, DOAEstimation_LowSNR}, the incorporation of a \ac{dnn} module can be computationally expansive to implement on some limited sensory devices. Moreover, while one can envision extensions to unsupervised operation (e.g., by leveraging structures in the signal model~\cite{weisser2022unsupervised} or in the augmented algorithm~\cite{revach2022unsupervised}) and to adaptivity across arrays (e.g., by on device learning~\cite{raviv2024adaptive} or using hypernetworks~\cite{ni2024adaptive}), in its current form, \acs{ssn} is trained in a supervised manner with sufficient data for a given array.}


The derivation of \acs{ssn} considers \ac{doa} recovery with \acp{ula}, as detailed in Subsection~\ref{ssec:Signal model}. However, its formulation can be extended to other forms of arrays where subspace methods are applied for \ac{doa} recovery, including planar arrays~\cite{chen2010esprit},  circular arrays~\cite{fuchs2001application}, and sparse arrays~\cite{guo2018doa}. For the latter, the ability of \acs{ssn} to learn to overcome array mismatches, as presented in Section~\ref{sec:Numerical Evaluation}, makes it extremely attractive. Furthermore, while we use \acl{rm} to train \acs{ssn}, one can also consider using \ac{esprit} during training, owing to its end-to-end differentiability. Specifically, since \ac{esprit} relies on signal subspace estimation (instead of noise subspace), using it during training can be beneficial when estimating a large number of sources. 
 {Moreover, we note that in addition to the traditional hyperparameters of deep learning models, \acs{ssn} introduces an additional hyperparameter, namely $\epsilon$ in \eqref{eqn:SurrogateCov}. In our numerical study in Section~\ref{sec:Numerical Evaluation} we keep its value fixed for all settings. Nonetheless, one can potentially perceive $\epsilon$ as being related to the noise variance in the representation of covariance matrix for narrowband signals in \eqref{covariance matrix equation}. This motivates exploring an \ac{snr}-dependent setting of $\epsilon$.}
We leave these extensions of \acs{ssn} for future study.  

%



%
 \vspace{-0.2cm}
\section{Numerical Evaluation}
\label{sec:Numerical Evaluation}
\vspace{-0.1cm}

In this section, we empirically evaluate \acs{ssn} when applied with different subspace and covariance-based estimation algorithms\footnote{The source code used in our empirical study along with the hyperparameters is available at \url{https://github.com/ShlezingerLab/SubspaceNet}}. Our experiments cover scenarios where assumptions \ref{itm:narrowband}-\ref{itm:Snapshots} are not necessarily satisfied, and demonstrate the ability of \acs{ssn} to enable classic \ac{doa} estimators to operate in these settings while preserving performance and interpretability. 
We first detail the experimental setup in Subsection~\ref{ssec:Experimental study}, after which we report the performances achieved under these setups in Subsection~\ref{ssec:Performances}. We conclude by investigating the interpretability of \acs{ssn} when combined with different \acl{mb} methods in Subsection~\ref{ssec:Interpretability}.

\vspace{-0.2cm}
\subsection{Experimental Setup}
\label{ssec:Experimental study}
\vspace{-0.1cm}

\subsubsection{Signal Model}
Throughout this study, we consider a \ac{ula} consisting of $N=8$ sensors with $M$ impinging signals, where we simulate different values of $M$. Except for the study reported in Subsection~\ref{sssec:Broadband settings}, where we consider broadband settings, the signals are generated according to \eqref{observation_formula}, i.e., representing a narrowband case. The \acp{doa} are uniformly generated from the interval $[-\frac{\pi}{2}, \frac{\pi}{2}]$. The \ac{snr} is defined as ${\rm SNR}=10\log_{10}\sigma_S^2/{\sigma_V^2}$, where $\sigma_S^2$ is the sources variance.

\subsubsection{\acs{ssn} Architecture} 
As detailed in Subsection~\ref{ssec:Architecture},  \acs{ssn} is comprised of a \ac{dnn} autoencoder with non-trainable preceding and subsequent processing. 
The auto-correlation features in \eqref{eqn:AutoCorr} are extracted for the value of the lag $\tau$, which is a hyper-parameter 
(set via manual tuning in our experiments),  that is bounded by the number of observations,  $\tau \le T-1$.
The autoencoder architecture utilized is composed of three convolution layers, having $16$, $32$, and $64$ output channels, respectively. These are followed by three deconvolution layers with $32$, $16$, and $1$ output channels, respectively. The kernel size of each layer is set to $2\times2$. The output of the autoencoder is reshaped into a surrogate covariance via~\eqref{eqn:SurrogateCov} with $\epsilon = 1$. The overall number of trainable parameters is $41,761$.
The architecture is trained using the Adam optimizer~\cite{kingma2014adam} with learning rate $\mu = 0.001$. For brevity, we use the abbreviated term {\em SubNet} for \acs{ssn} in the legends of the figures.

\subsubsection{\ac{doa} Estimators}
\label{sssec:doaBench}
To assess the ability of \acs{ssn} to improve classic subspace methods, we evaluate the classic subspace methods detailed in Subsection~\ref{ssec:subspace_methods}, i.e., \ac{music}, \ac{rm}, and \ac{esprit}, comparing their performance when combined with \acs{ssn} to that achieved when applied with the standard empirical covariance.  Specifically, when employing \ac{music}, we detect the \acp{doa} from the spectrum by searching for the peaks over a grid with resolution  $0.01^{\circ}$.

In settings where  \ref{itm:narrowband}-\ref{itm:Snapshots} do not hold, we contrast \acs{ssn} with representative \acl{mb} extensions of classic subspace methods designed to cope with each scenario. In particular,  for coherent sources, we employ \ac{sps} pre-processing, which is a widely adopted \acl{mb} approach for coping with such scenarios \cite{wang1994spatial, 1518905}. 
 {When considering narrowband non-coherent sources, we also compare the \ac{doa} estimators with the corresponding \ac{zzb} derived in \cite{zhang2022ziv}.}
To evaluate \ac{doa} estimation under broadband settings,  we compare \acs{ssn} with \acl{mb} extensions for broadband signals based on the technique proposed in \cite{yoon2006tops}. This involves dividing the relevant frequency spectrum into bins, and then applying classical narrowband subspace methods to each bin separately, while aggregating the results to yield a broadband estimate. 

We also compare \acs{ssn} to alternative \ac{dnn}-based \ac{doa} estimators. We consider two data-driven benchmarks: $1)$ a \ac{cnn}-based \ac{doa} estimator based on the architecture proposed in \cite{DOAEstimation_LowSNR} coined {\em CNN}, representing a black-box \ac{dnn} design; $2)$ {\em DA-MUSIC} of \cite{merkofer2022deep}, which interconnects a \acl{rnn} and a fully-connected architecture following the flow of \ac{music}. All data-driven \ac{doa} estimators are trained on the same data, comprised of $J=45000$ samples corresponding to \ac{snr} of $10$ dB (unless stated otherwise).  {When evaluating the data-driven estimators with varying number of sources, the data is comprised of all considered values of $M$, and training is followed by adaption to the considered value of $M$ using $8000$ samples with $M$ sources.} 
For the {\em CNN}  model, a grid of classes with a resolution of $0.5^{\circ}$ is defined, as in the original paper, resulting in over $21\cdot 10^6$ trainable parameters (higher resolution results in a substantially large and computationally intensive number of trainable parameters), while DA-MUSIC has $10,194$ trainable parameters. {Following the approach used in \cite{DOAEstimation_LowSNR}, a dataset containing ${M \choose 361}$ combinations of angles is generated for each tested \ac{snr}, resulting in an extensive set of over $300,000$ samples for training the {\em CNN} model}. 

\vspace{-0.2cm}
\subsection{\ac{doa} Recovery Performance}
\label{ssec:Performances}
\vspace{-0.1cm}
Here, we evaluate the accuracy of \acs{ssn}, measured via the  \ac{rmspe}. We consider settings with coherent sources (not holding \ref{itm:coherent}); with dominant noise and limited snapshots (not holding \ref{itm:Snapshots}); with broadband signals (not holding \ref{itm:narrowband}); and with miscalibrated arrays (not holding \ref{itm:callib}). We empirically evaluate the \ac{rmspe} by averaging over $5000$ Monte Carlo trials. To focus on the ability to identify the \acp{doa} correctly, here we set $\hat{M}$ to the true number of \acp{doa} $M$, while in Subsection~\ref{ssec:Interpretability} we show that $\hat{M}$ can indeed be extracted as an accurate estimate of $M$. For ease of presentation, figures comparing \ac{doa} estimators versus \ac{snr} report performance in terms of mean squared periodic error in dB.

\subsubsection{Coherent Sources}
\label{sssec:Coherent Sources}
We first evaluate the ability of \acs{ssn} to effectively handle coherent sources. We focus on the challenging fully coherent case, where  all sources exhibit identical phases and amplitudes, resulting in $\myMat{R}_{S}$ having minimal unit rank. We simulate different numbers of narrowband coherent sources estimated from $T=100$ snapshots, with an \ac{snr} of $10$ dB and a calibrated \ac{ula}.

The \ac{rmspe} results achieved by the considered \ac{doa} estimators are reported in Table~\ref{table:coherent sources sim}. We observe in Table~\ref{table:coherent sources sim} that \acs{ssn} augmentation enables effective and consistent handling of coherent sources, improving the performance of all considered \acl{mb} algorithms. Notably, the augmentation of \ac{rm} with \acs{ssn} on $M=2$ sources leads to almost $60\times$ better estimation performance than classical \ac{rm}. Even when a dedicated \ac{sps} technique is used to increase the empirical covariance rank, the \ac{sps} \ac{rm} algorithm results are inferior to \acs{ssn}. We also observe that while \acs{ssn} was trained with \ac{rm}, its surrogate covariance also notably facilitates the operation of  \ac{music} and \ac{esprit}, which both exhibit improvement in estimation performance compared to classical and pre-processed estimation methods.
Moreover, \acs{ssn} outperforms the data-driven benchmarks, while using much less parameters than {\em CNN} model and maintaining a similar scale of parameter quantity as {\em DA-MUSIC}, without limiting the interpretability of \acl{mb} \ac{doa} estimators, as shown in Subsection~\ref{ssec:Interpretability}.

%
\begin{table}
\begin{center}
{
\vspace{0.3cm}
\resizebox{0.95\columnwidth}{!}{
\begin{tabular}{|l|c|c|c|}
\hline
\rowcolor{Gray}
\text{RMSPE [$^\circ$]}& \text{$M = 2$} & \text{$M = 3$} & \text{$M = 4$} \\ \hline
\textit{\ac{rm}}      & 12.4790& 20.3972& 23.2047
\\ \hline
\textit{SPS-\ac{rm}}  & 1.0112& 1.6369& 12.1467
\\ \hline
\rowcolor{LightCyan}
\textit{\acs{ssn}+\ac{rm}}  & \textbf{0.2005}& \textbf{0.7219}& \textbf{3.8846}\\ \hline
\textit{\ac{music}}           & 9.4709& 19.0794& 22.8037
\\ \hline
\textit{SPS-\ac{music}}       & 2.5840& 2.3342& 18.3323
\\ \hline
\rowcolor{LightCyan}
\textit{\acs{ssn}+\ac{music}} & 2.1084& 4.2398& 5.2482
\\ \hline
\textit{\ac{esprit}}      & 25.8426& 22.4599& 23.2047
\\ \hline
\textit{SPS-\ac{esprit}}  & 0.9224& 3.7184& 12.5477
\\ \hline
\rowcolor{LightCyan}
\textit{\acs{ssn}+\ac{esprit}} & 0.4761& 2.4350& 4.4690
\\ \hline
\textit{DA-MUSIC}      & 1.1860& 2.6585& 3.9705
\\ \hline
\textit{CNN}  & 0.7693& 1.2438& 7.4198\\ \hline
\end{tabular}
}
\caption{\ac{doa} accuracy, coherent sources.}
\label{table:coherent sources sim}
}
\vspace{-0.3cm}
\end{center}
\end{table}

\

\subsubsection{Low SNR and Few Snapshots}
\label{sssec:snapshots & SNR}
We proceed to evaluate the ability of \acs{ssn} to facilitate coping with low \acp{snr} and a limited number of snapshots. Such conditions are recognized to profoundly deteriorate \acl{mb} \ac{doa} estimation. To pinpoint the gains of \acs{ssn}, we first consider only low \acp{snr} with $M=2$ non-coherent sources, after which we limit the number of snapshots, and then replace the sources  with coherent ones, while systematically contrasting \acs{ssn}-aided \ac{doa} estimators with the classic model-based operation for each scenario.  In experiments conducted with multiple \acp{snr}, the training data used by data-driven estimators is that of the lowest \ac{snr} considered in the range.

We thus commence with considering highly noisy observations with \acp{snr} in the range of $[-5,0]$ dB. The signals are non-coherent and the number of snapshots is $T=200$.  {Since \ac{sps} is not designed to enhance subpsace methods in such scenarios with non-coherent sources and sufficient snapshots, we do not include \ac{sps}-based subspace methods here. Accordingly, we focus on comparing their operation with the conventional empirical covariance to that provided by \acs{ssn}}.
The results, reported in Fig. \ref{fig:DoA estimation_$T=200$, non-coherent}, demonstrate that \acs{ssn} augmentations notably improve the ability of subspace-based \ac{doa} estimators in coping with high noise levels. These results emphasize that \acs{ssn} can not only mitigate the harmful effects of coherency, but also effectively enhance the \ac{snr}, leading to improved accuracy in \ac{doa} estimation,  {while reaching performance within a limited gap from the lower bound on the achievable accuracy dictated by the \ac{zzb}.}

%
\begin{figure}
\begin{center}
\includegraphics[width=\columnwidth]{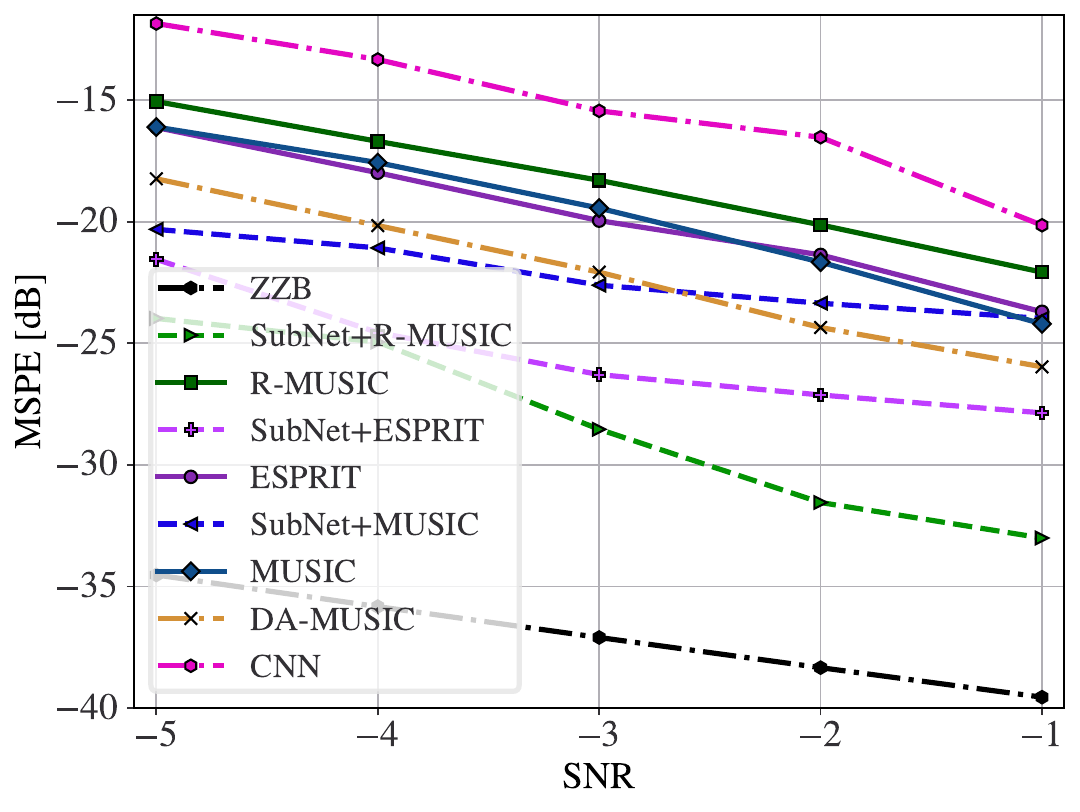}
\caption{\ac{doa} estimation MSPE, $T=200$, non-coherent sources}%
\label{fig:DoA estimation_$T=200$, non-coherent}
\end{center}
\end{figure}

We proceed to considering the challenging setting where merely $T=2$ snapshots are employed. Here, we set the \ac{snr} to be in the range of $[5, 10]$ dB. The resulting \ac{doa} estimation accuracy, depicted in Fig.~\ref{fig:DoA estimation_$T=2$, non-coherent}, clearly shows that the incorporation of  \acs{ssn}  into classical \acl{mb} methods enables them to effectively handle few observations, even in a scenarios where these methods are typically unfeasible, i.e., where the number of snapshots is not larger than the number of sources. Moreover, the performance of \acs{ssn} continues to improve as the  \ac{snr} increases, while the estimation accuracy of classical \acl{mb} methods is dominated by their inability to cope with few snapshots, and thus does not notably improve with \ac{snr}.

%
\begin{figure}
\begin{center}
\includegraphics[width=\columnwidth]{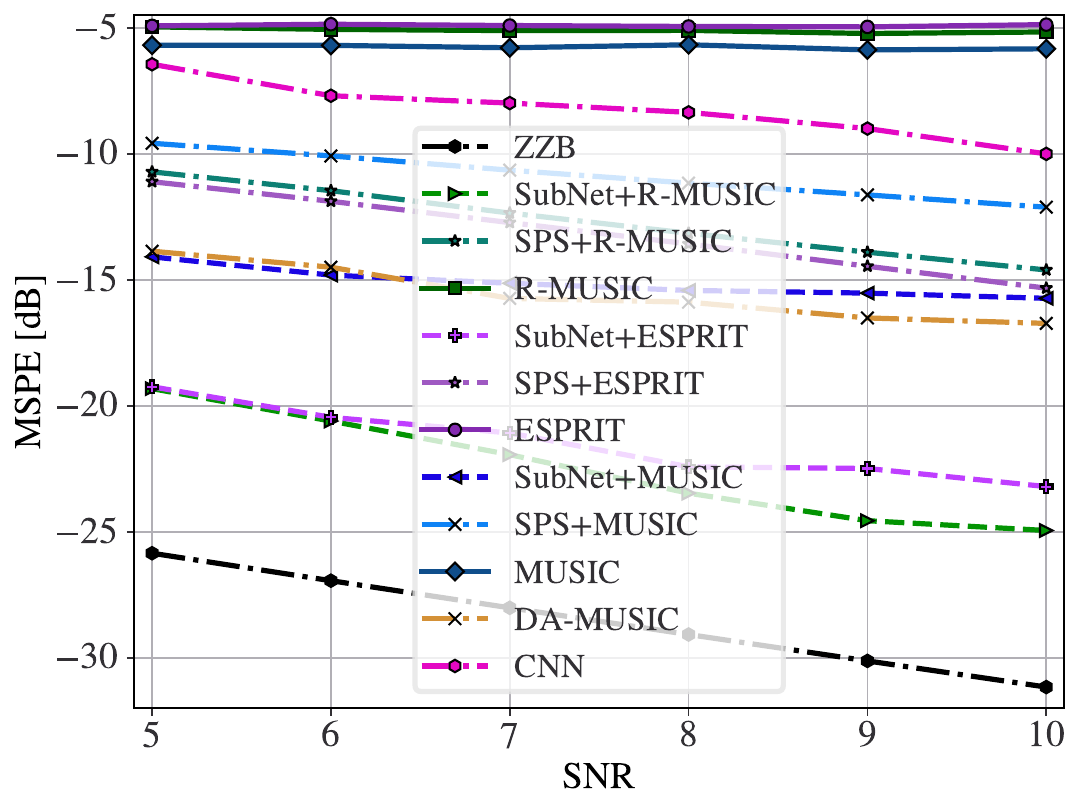}
\caption{\ac{doa} estimation MSPE, $T=2$, non-coherent sources}%
\label{fig:DoA estimation_$T=2$, non-coherent}
\end{center}
\end{figure}
\begin{figure*}
\begin{center}
\begin{subfigure}[pt]{0.32\linewidth}
\includegraphics[width=\linewidth]{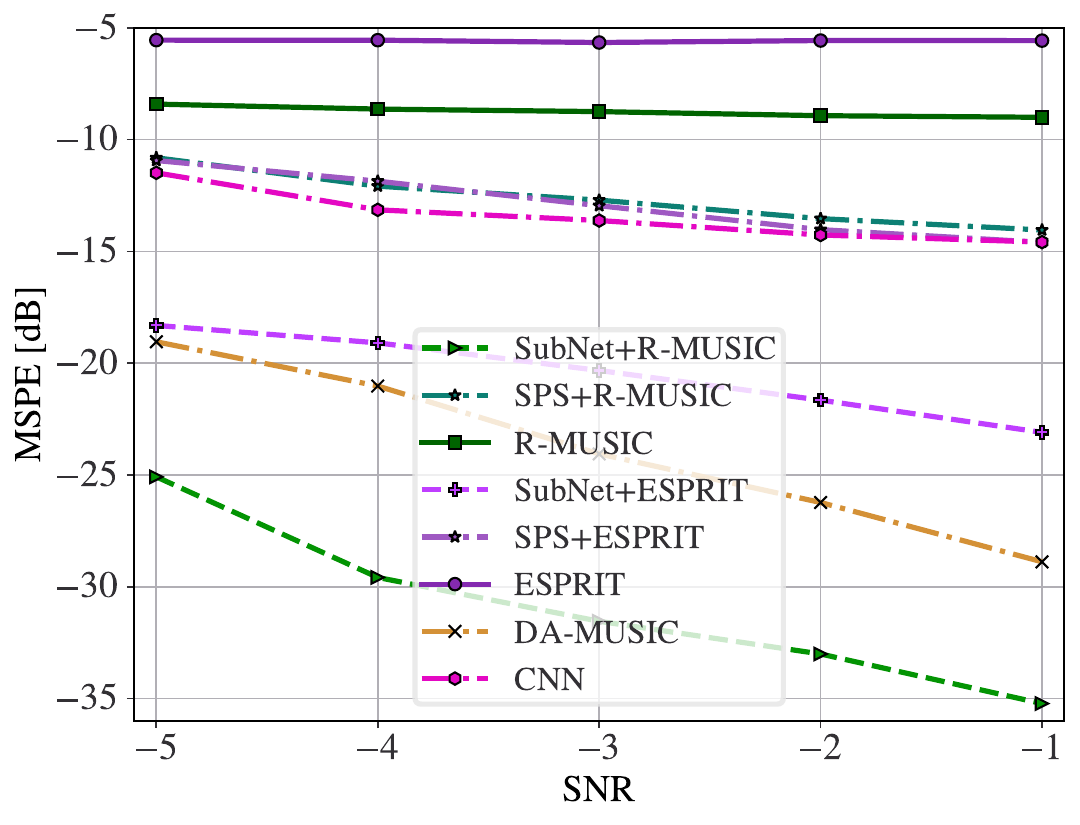}
\caption{$T=200$, low \ac{snr}}
\label{fig:SNR_coherent_$T=200$}
\end{subfigure}
\begin{subfigure}[pt]{0.32\linewidth}
\includegraphics[width=\linewidth]{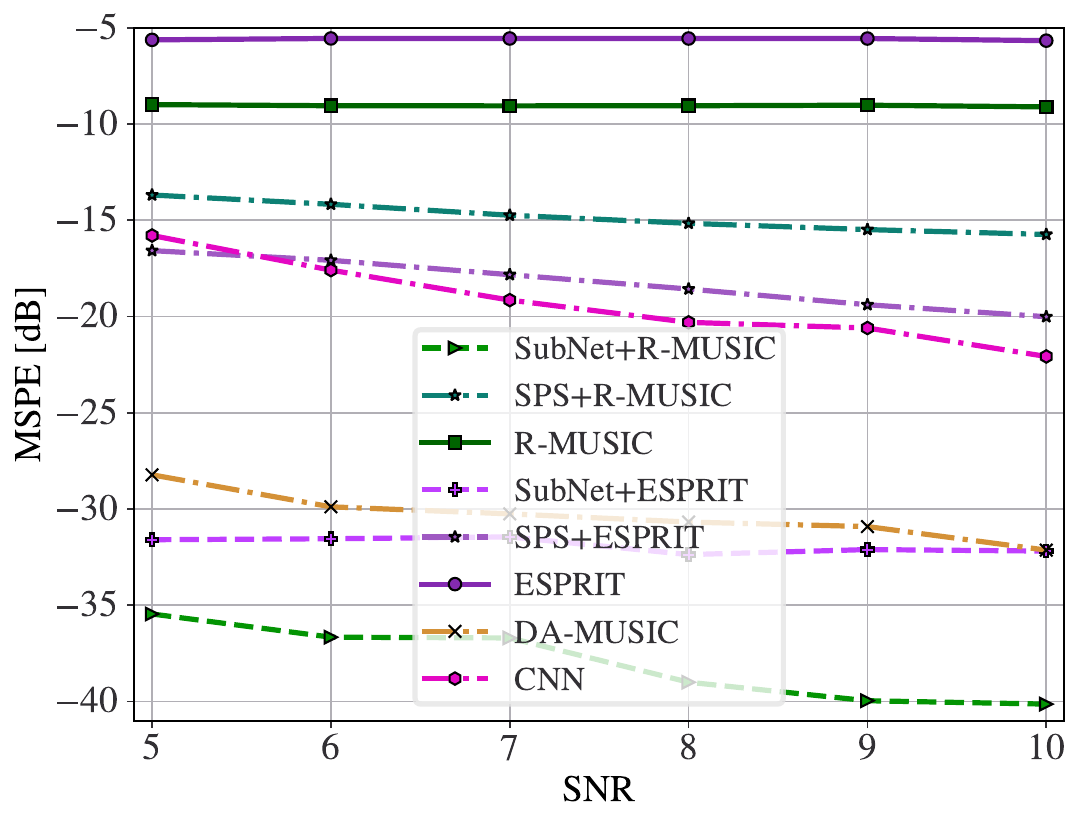}
\caption{$T=20$, moderate \ac{snr}}
\label{fig:SNR_coherent_$T=20$}
\end{subfigure}
\begin{subfigure}[pt]{0.32\linewidth}
\includegraphics[width=\linewidth]{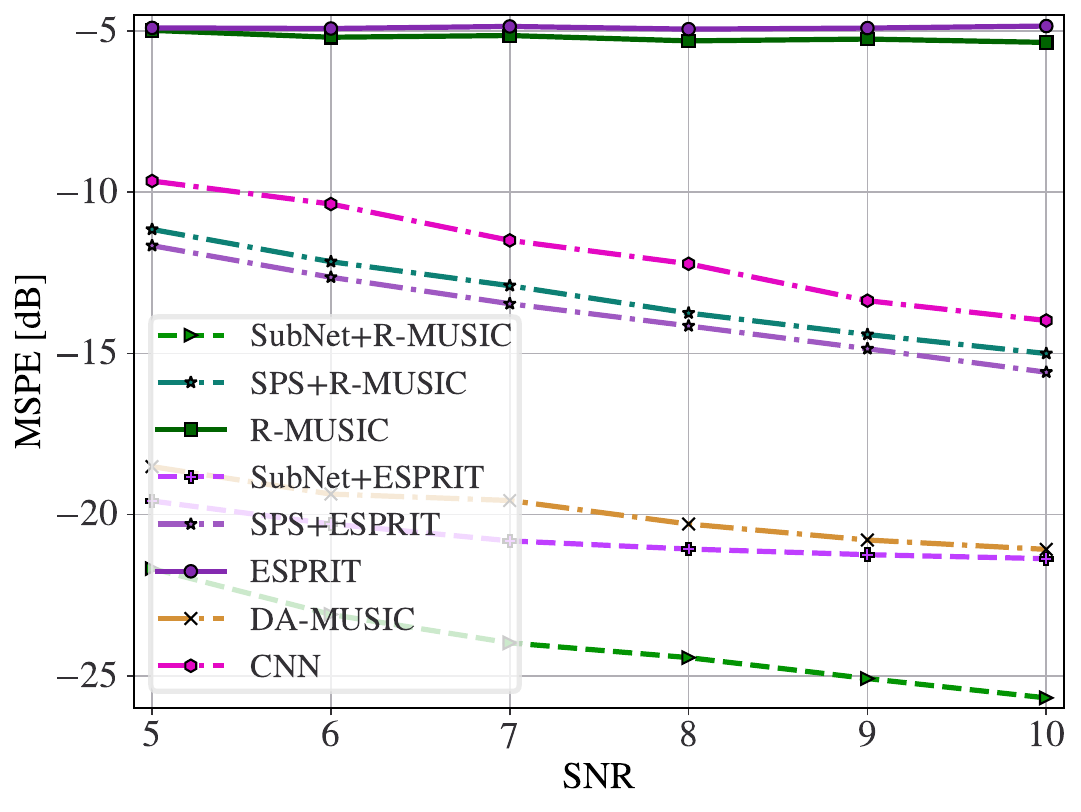}
\caption{$T=2$, moderate \ac{snr}}
\label{fig:SNR_coherent_$T=2$}
\end{subfigure}
\caption{\ac{doa} estimation, SNR $\&$ minor snapshots, coherent sources}%
\label{fig:DoA estimation SNR, snapshots and coherent sources}
\end{center}
\end{figure*}

%
We next introduce coherent sources, assessing the ability of \acs{ssn} in handling settings not meeting both \ref{itm:coherent} and \ref{itm:Snapshots}. We compare the augmentation of  subspace methods with \acs{ssn} to their \acl{mb} counterparts in Fig.~\ref{fig:DoA estimation SNR, snapshots and coherent sources} under low \acp{snr} with $T=200$ snapshots (Fig.~\ref{fig:SNR_coherent_$T=200$}) and under moderate \acp{snr} with merely $T=20$ and $T=2$ snapshots (Figs.~\ref{fig:SNR_coherent_$T=20$}-\ref{fig:SNR_coherent_$T=2$}, respectively).  {For brevity and to avoid cluttering, we include two subspace methods in these figures -- \ac{rm} and \ac{esprit} -- applying each with the conventional empirical covariance, with \ac{sps}, and with \acs{ssn}.} The results show that the surrogate covariance of \acs{ssn} enables reliable subspace-based \ac{doa} estimation even in an extreme scenario where  \ref{itm:coherent} and \ref{itm:Snapshots} are both violated. This is also evident in Table~\ref{table:extreme scenario sim}, which reports the  \ac{rmspe} values achieved for SNR of $5$ dB, comparing \acs{ssn} also to the data-driven benchmarks, whose performance is inferior to that of \acs{ssn}.

%

\begin{table}
\begin{center}
{
\vspace{0.3cm}
\resizebox{0.65\columnwidth}{!}{%
\begin{tabular}{|l|l|}
\hline
\rowcolor{Gray}
\text{Algorithm}                & \text{RMSPE [$^\circ$]}\\ \hline
\textit{\ac{music}}                  & 25.8134
\\ \hline
\textit{\ac{sps}+\ac{music}}& 8.0758
\\ \hline
\rowcolor{LightCyan}
\textit{\acs{ssn}+\ac{music}}           & \textbf{2.2532}
\\ \hline
\textit{\ac{esprit}}                 & 29.0632
\\ \hline
\textit{\ac{sps}+\ac{esprit}}                 & 11.0179
\\ \hline
\rowcolor{LightCyan}
\textit{\acs{ssn}+\ac{esprit}}          & \textbf{1.9675}
\\ \hline
\textit{\ac{rm}}                 & 26.3560
\\ \hline
\textit{\ac{sps}+\ac{rm}}                 & 8.2735
\\ \hline
\rowcolor{LightCyan}
\textit{\acs{ssn}+\ac{rm}}          & \textbf{1.8907}
\\ \hline
\textit{DA-MUSIC}          & 4.7326
\\ \hline
\textit{CNN}          & 9.6053
\\ \hline
\end{tabular}
}
\caption{$M=2$ coherent signals, $T=2$, SNR=$5$ dB}
\label{table:extreme scenario sim}
}
\vspace{0.3cm}
\end{center}
\end{table}
%

\begin{figure}
\begin{center}
\includegraphics[width=\columnwidth]{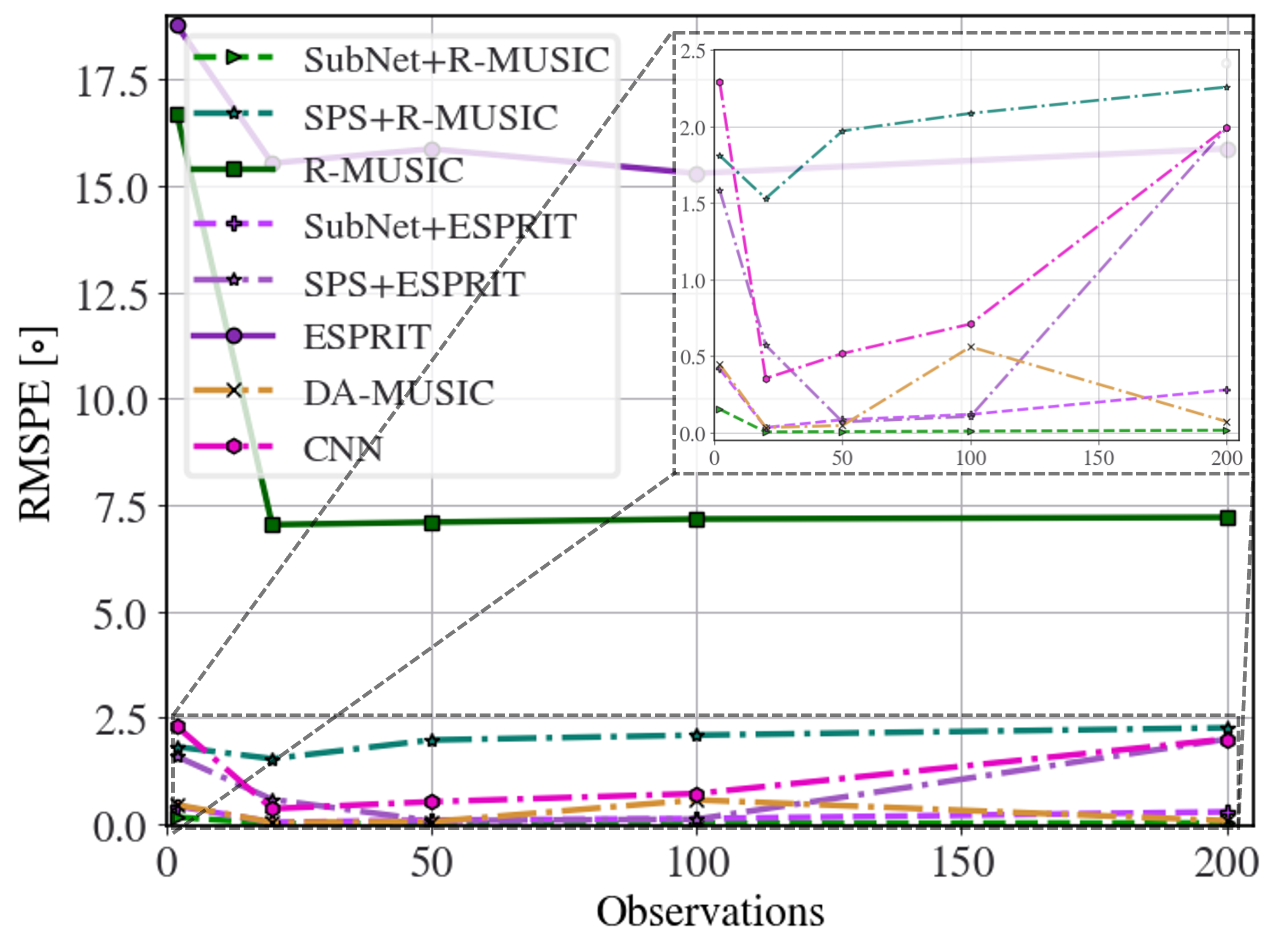}
\caption{\ac{doa} estimation accuracy versus $T$, $M=2$ coherent sources, moderate \ac{snr}}%
\label{fig:multiple_coherent_all_multiple_zoom_out}
\end{center}
\end{figure}

To further showcase the ability of \acs{ssn} to enable subspace methods to cope with limited snapshots, we evaluate the \ac{rmspe} of the considered methods for different number of snapshots. As in Figs.~\ref{fig:SNR_coherent_$T=20$}-\ref{fig:SNR_coherent_$T=2$}, we consider $M=2$ coherent sources at \ac{snr} of $5$ dB, and report the performance of \ac{rm} and \ac{esprit} with and without \acs{ssn} compared to the data-driven benchmarks in Fig.~\ref{fig:multiple_coherent_all_multiple_zoom_out}. We observe in Fig.~\ref{fig:multiple_coherent_all_multiple_zoom_out} that subspace methods combined with \acs{ssn} consistently achieve improved accuracy, and  that \acs{ssn}  allows subspace methods to operate reliably with limited snapshots.  Conventional covariance processing methods reach an error floor, and possibly even degrade in some cases when combining more snapshots representing coherent sources.

%

\color{black}
\subsubsection{Broadband settings}
\label{sssec:Broadband settings}
The scenarios considered so far were all based on the narrowband formulation in~\eqref{observation_formula}. 
Here, we evaluate \acs{ssn} under broadband settings. In such cases, one can formulate the signal model in the frequency domain, which at frequency $\omega$ is given by~\cite{yoon2006tops}
\begin{equation}\label{observation_BB_formula}
  \ft{X}(\omega) = \myMat{A}(\omega, \myVec{\theta}) \ft{S}(\omega) + \ft{V}(\omega).
\end{equation}
In \eqref{observation_formula}, $ \ft{X}(\omega) $ and $ \ft{S}(\omega) $ are the discrete-time multivariate Fourier transforms of $\myVec{x}(t)$ and $\myVec{s}(t)$, respectively, $\ft{V}(\omega)$ is noise, and the steering vectors comprising $\myMat{A}(\omega, \myVec{\theta})$ are 
\begin{equation}
    \myVec{a}(\omega,\theta) = \begin{bmatrix}1, \ e^{- j \omega \frac{d}{c} \sin{\theta}}, \ \dots, \ e^{- j \omega (M-1) \frac{d}{c}\sin{\theta}}\end{bmatrix},
\end{equation}
with $d$ denoting the element spacing of the \ac{ula}, and $c$ being the propagation velocity.


We simulate the time-domain broadband signals using \ac{ofdm} with $L$ sub-carriers, where each carrier. denoted $ \myScal{s}_{m,l}$ for the $l$th subcarrier of the $m$th signal, is modulated independently with a zero-mean complex-Gaussian unit variance distribution. The overall \ac{ofdm} signal can be expressed as:
\begin{equation}
\label{eqn:OFDM_signal_model}
    \myScal{s}^{\rm (OFDM)}_{m}(t) = \frac{1}{L} \sum_{l=0}^{L-1} \myScal{s}_{m,l} e^{2\pi j l\frac{B_{\rm f}  }{L f_{\rm s}}t}
\end{equation}
where $f_{\rm s}$ is the sampling frequency, and $B_{\rm f}$ is the signal bandwidth.
The acquired observations are generated following \eqref{observation_BB_formula}, where the inter-element spacing is given by half the minimal wavelength $d = c / 2 B_{\rm f}$. 
We generate $M=2$ \ac{ofdm} signals with $L=500$ subcarriers, setting  $B_f = 0.5$ kHz. 
Various sampling rates ranging from $50$ to $1000$ Hz are tested, while the signals are observed for a duration of $1$ second, thus the number of snapshots is $T = 1/f_{\rm s}$. The performance of  \acs{ssn} is evaluated for all sampling rates, despite being trained solely on observations sampled at $0.2$ kHz.

\begin{figure*}
\begin{center}
\begin{subfigure}[pt]{0.49\linewidth}
\centering
\includegraphics[width=\columnwidth]{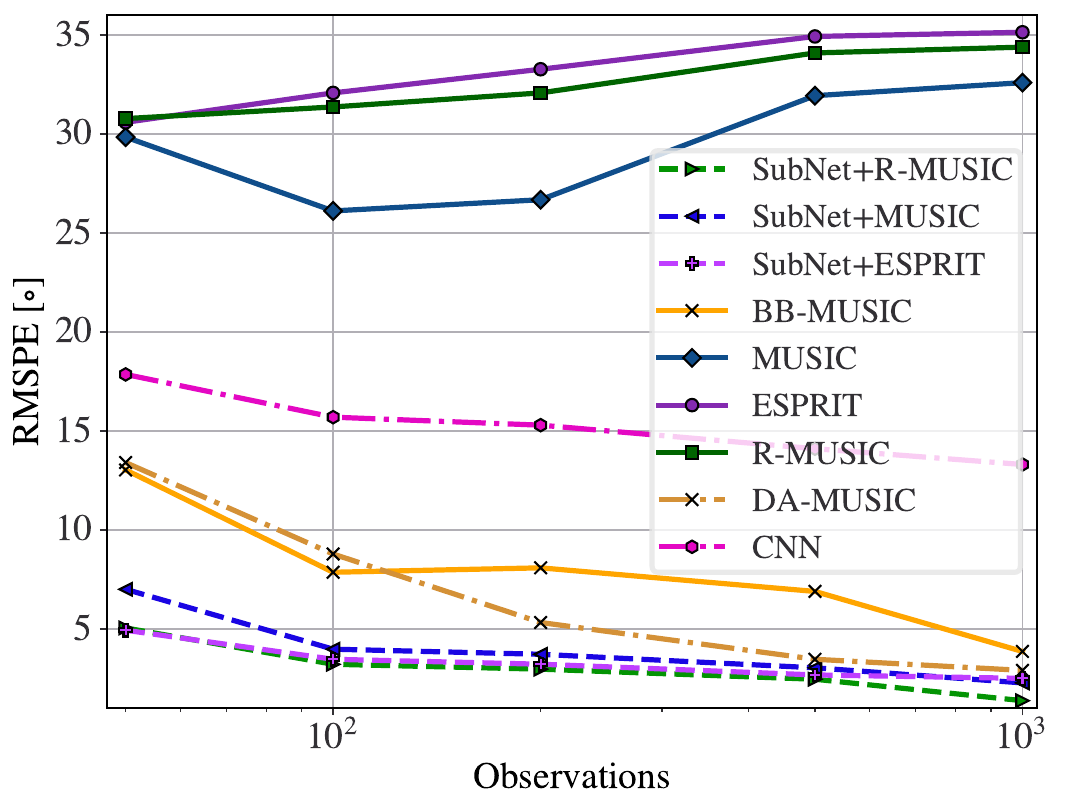}
\caption{Non-coherent OFDM signals}
\vspace{0.2cm}
\label{fig:OFDM_non_coherent}
\end{subfigure}
\begin{subfigure}[pt]{0.49\linewidth}
\centering
\includegraphics[width=\columnwidth]{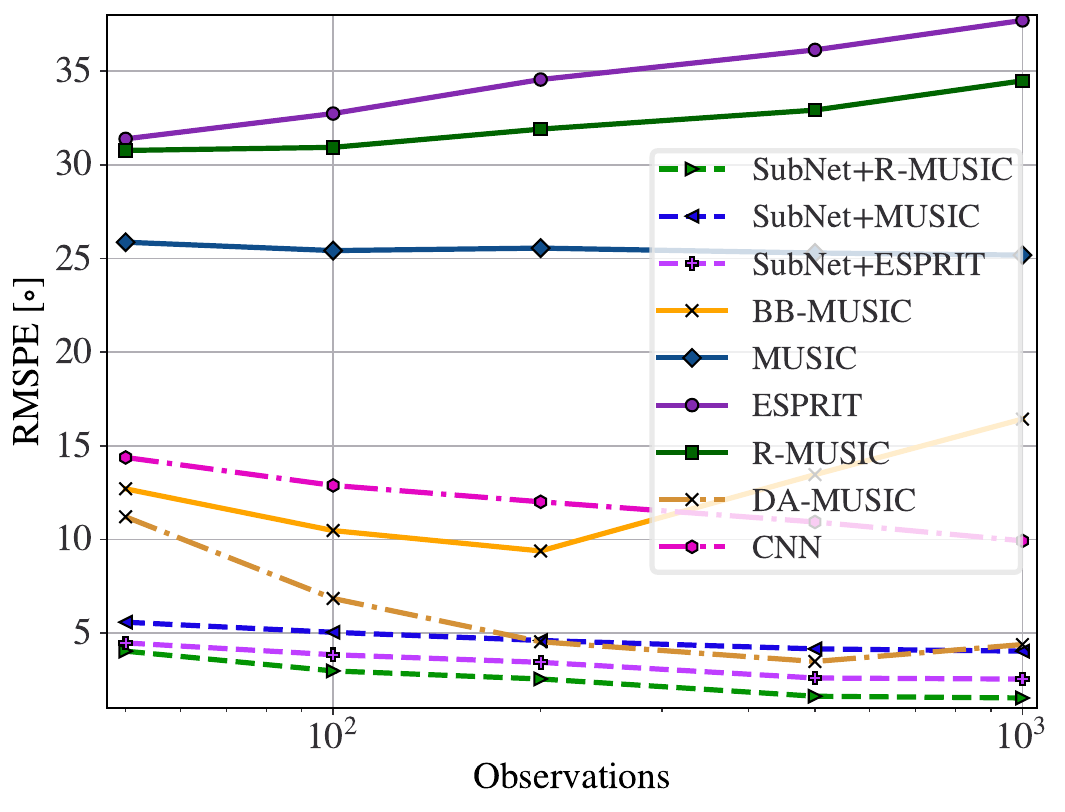}
\caption{Coherent OFDM signals}
\vspace{0.2cm}
\label{fig:OFDM_coherent}
\end{subfigure}
\caption{Broadband \ac{doa} estimation comparison, for various observations}
\label{fig:OFDM_comparisom}
\end{center}
\vspace{-0.4cm}
\end{figure*}

We evaluate the performance of \ac{music}, \ac{rm}, and \ac{esprit} when combined with \acs{ssn} and when using the conventional empirical covariance, as well as compared to the broadband extension of \ac{music} proposed in \cite{yoon2006tops} which divides the bandwidth into $50$ separate bins.  The results, depicted in Fig.~\ref{fig:OFDM_comparisom}, consider both non-coherent  as well as coherent sources. 
The non-coherent results depicted in Fig.~\ref{fig:OFDM_non_coherent} demonstrate that the incorporation of \acs{ssn} models leads to significantly improved performance in comparison to narrowband methods and the broadband extension of \ac{music}.  \acs{ssn}  consistently enables reliable \ac{doa} estimation, even when a minimal number of snapshots are available, with $f_{\rm s}=50$ Hz. These findings also hold when the sources are coherent, as demonstrated in Fig.~\ref{fig:OFDM_coherent}. In contrast, as expected, the subspace \acl{mb} broadband extension is unable to estimate \acp{doa} under both coherent and broadband conditions. 

To numerically capture the exact gains of \acs{ssn}, we report the \ac{rmspe} values achieved by \acs{ssn} in the broadband setting with $M=2$ coherent sources and a sampling rate of $f_{\rm s} = 200Hz$ in Table~\ref{table: Broadband OFDM signals sim}. We also include the accuracy achieved by the data-driven DA-MUSIC and CNN benchmarks. Table \ref{table: Broadband OFDM signals sim} showcases that \acs{ssn} allows model-based subspace methods to achieve substantially improved accuracy in broadband settings, surpassing the data-driven benchmarks, as well as model-based methods designed for broadband settings, whose inference procedure is more complex as they require separate \ac{doa} estimation for each bin. 


%
%



\begin{table}
\begin{center}
{
\resizebox{0.65\columnwidth}{!}{%
\begin{tabular}{|l|l|}
\hline
\rowcolor{Gray}
\text{Algorithm}                & \text{RMSPE [$^\circ$]}\\ \hline
\textit{\ac{music}}                  & 26.7513\\ \hline
\textit{Broadband \ac{music}}        & 7.06456\\ \hline
\textit{DA MUSIC}   & 4.74523\\ \hline
\textit{CNN}   & 26.7439\\ \hline
\rowcolor{LightCyan}
\textit{\acs{ssn}+\ac{music}}           & \textbf{2.2861}\\ \hline
\textit{\acs{ssn}+\ac{esprit}}          & 3.7127\\ \hline
\textit{\acs{ssn}+\ac{rm}}               & 2.3376\\ \hline
\end{tabular}
}
\caption{\ac{rmspe} results, $M=2$ broadband OFDM signals}
\label{table: Broadband OFDM signals sim}
}
\end{center}
\end{table}
%




\subsubsection{Array Miscalibration}
\label{sssec:miscalibration}
Here, we investigate the effectiveness of \acs{ssn} in learning from data to handle array miscalibrations.
The calibration of the array manifold is a critical aspect of \ac{doa} estimation, as inaccurate calibration can lead to substantial errors.
We simulate two forms of miscalibrations, which both cause the actual array response to deviate from the ideal  pattern, significantly affecting the performance of subspace methods. In both scenarios, the received signal is comprised of $M=2$  non-coherent signals. 

In the first scenario, array distance miscalibration is simulated, where the spacing between each pair of sensors deviates from the standard \ac{ula} configuration of half-wavelength, resulting in non-uniform element spacing. The spacing between each adjacent pair of sensors in the $m$th position is established as the nominal half-wavelength distance $d$ plus a uniformly distributed random variable denoted as $\delta_m~\sim U(-\eta, \eta)$, where $\eta$ is the percentage of deviation from the nominal spacing. Consequently, the  steering vector  in \eqref{eqn:SteeringVec} becomes miscalibrated, with it $m$th element now given by
\begin{equation}
\label{eqn:miscalibrated SteeringVec}
    [\myVec{a}(\theta)]_m \triangleq e^{-2\pi j \frac{(d + \delta _m)(m-1)}{c}  \sin(\theta)}, \quad m\in\{1,\ldots,M\}.
\end{equation} 

\begin{figure*}
\begin{center}
\begin{subfigure}[pt]{0.49\linewidth}
\centering
\includegraphics[width=\columnwidth]{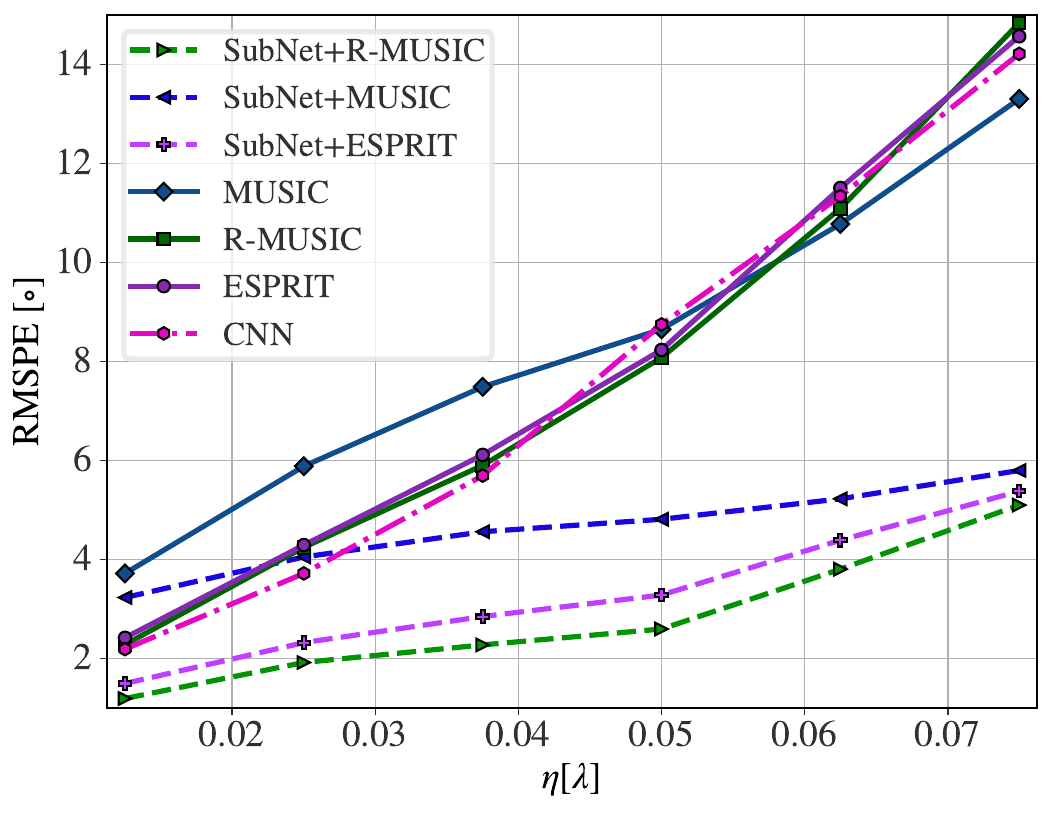}
\caption{Mismatch due to deviation from half-wavelength spacing}.
\label{fig:distance_calibration}
\end{subfigure}
\begin{subfigure}[pt]{0.49\linewidth}
\centering
\includegraphics[width=\columnwidth]{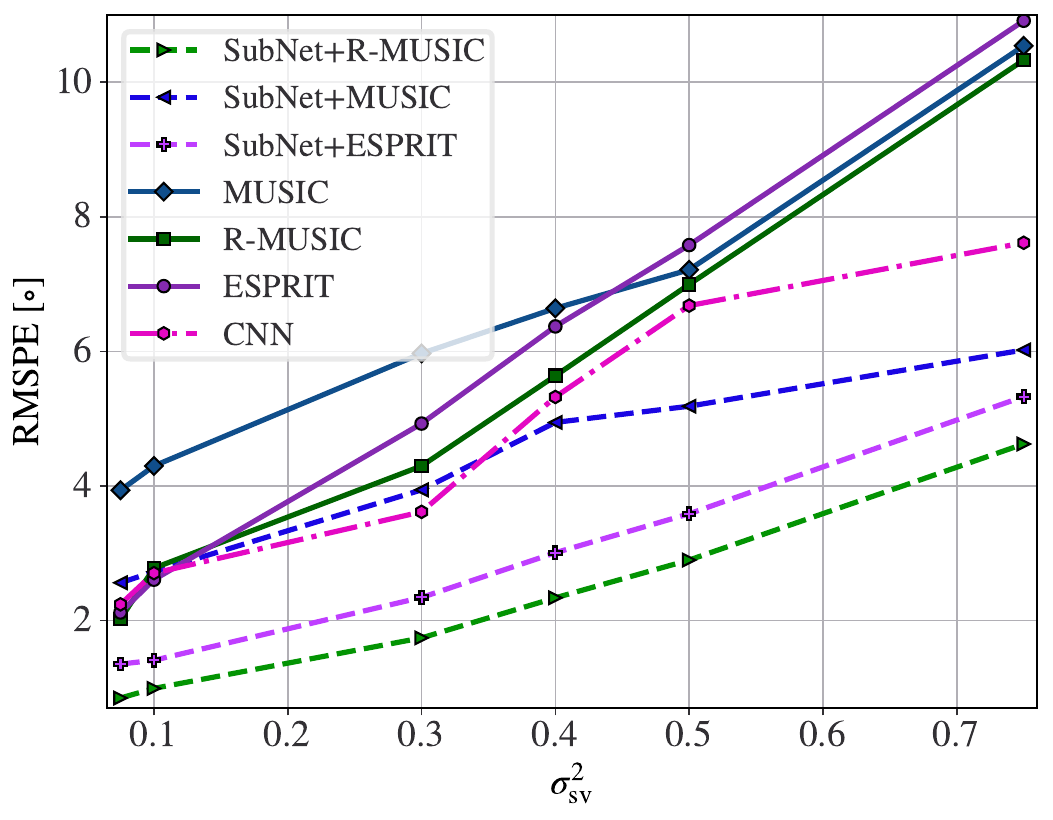}
\caption{Mismatch due to noisy steering vectors}
\label{fig:steering_vector_noise}
\end{subfigure}
\caption{Miscalibrated array \ac{doa} estimation comparison}
\label{fig:miscalibration}
\vspace{-0.4cm}
\end{center}
\end{figure*}

The resulting \ac{rmspe} values achieved by subspace-based \ac{doa} estimators with and without \acs{ssn} are reported in Fig.~\ref{fig:distance_calibration}, for different values of $\eta \in [0.025d, 0.15d]$, i.e., a maximum deviation of  $30\%$ from the calibrated spacing. The results depicted in Fig.~\ref{fig:distance_calibration} underscore the ability of \acs{ssn} to deal with discrepancies in the array geometry, particularly when the steering vector is affected by non-uniform (and non-nominal) elements spacing. In contrast, classical subspace methods heavily rely on the nominal \ac{ula} form of the array manifold,  yielding non-optimal estimation as $\eta$ increases.

In the second scenario, we introduce miscalibration directly in the steering vector, rather than in the element spacing. This is achieved by adding a zero-mean complex-Gaussian noise with variance $\sigma_{\rm sv}^2$ to each element in the  steering vector, resulting in mismatches in the creation of the subspace core equation  \eqref{basic_core_equation}. The simulation results depicted in Fig.~\ref{fig:steering_vector_noise} further indicate that  \acs{ssn} indeed exhibits superior performance over traditional subspace methods when handling steering vector corruption due to the addition of zero-mean Gaussian noise, resulting in more accurate \ac{doa} estimation. 
On the other hand, classical subspace methods appear to be highly sensitive to this type of corruption, leading to poor estimation results as the noise variance increases. Additionally, the results indicate that the performance gap between traditional subspace methods and \acs{ssn} increases as the level of steering vector corruption increases, for both $\sigma_{sv}^2 = 0.75$ and $\eta = 0.075d$. This suggests that \acs{ssn} could potentially provide more significant benefits in scenarios with higher levels of noise corruption and geometry mismatches, making it a promising solution for applications in challenging environments.


\vspace{-0.2cm}
\subsection{Interpretability}
\label{ssec:Interpretability}
\vspace{-0.1cm}
The results presented so far systematically demonstrate that the augmentation of subspace methods with \acs{ssn} enables these classic algorithms to operate reliably in scenarios where their traditionally limiting assumptions \ref{itm:narrowband}-\ref{itm:Snapshots} do not hold. 
Specifically, \acs{ssn} outputs a surrogate covariance matrix which enables subspace methods to operate in an unaltered manner in various challenging scenarios: It can efficient cope with wideband signals; accurately resolve  coherent signals; operate reliably with few snapshots and high noise levels; and boost robustness to different forms of miscallibrations. These gains stem from the fact that \acs{ssn} leverages the abstractness of \acp{dnn} to learn a desirable mapping from data, which here is the extraction of a suitable covariance matrix. An additional key gain of \acs{ssn}, which is not shared by conventional \ac{dnn}-based solutions, is its ability to preserve the desirable interpretable operation of model-based \ac{doa} estimators, and in fact provide meaningful visual representations associated with classic methods in settings where these are typically not achievable. To show this property of \acs{ssn}, we first visualize its ability to yield distinguishable signal and noise subspaces. Then we show the spectrum representations obtained by \ac{music} and \ac{rm} when combined with \acs{ssn}, after which we demonstrate that \acs{ssn} can yield meaningful beampatterns when combined with covariance-based beamformers that are not necessarily based on subspace methods, such as \ac{mvdr}.  


\subsubsection{Subspace Separation}
\label{sssec:Subspace separation}

\begin{figure}
\centering
\includegraphics[width=\columnwidth]{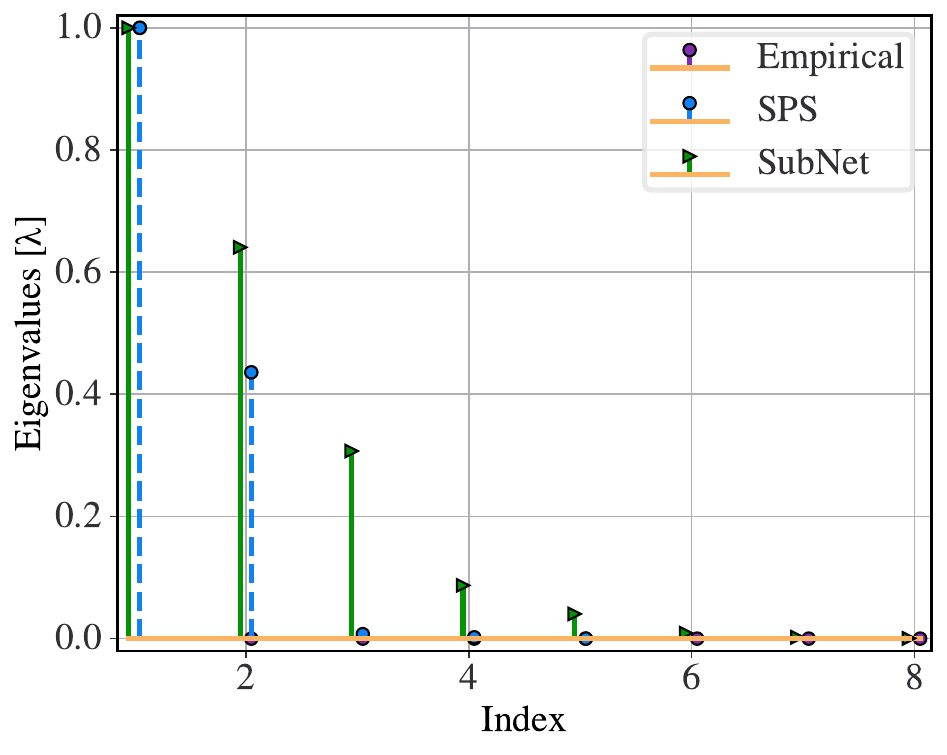}
\caption{Normalized eigenvalues of empirical, \ac{sps} and \acs{ssn} covariance matrices. $M=3$ coherent sources.}
\label{fig:eigenvalues}
\end{figure}

We commence with a comparison of the normalized covariance eigenvalues produced by \acs{ssn} compared with the conventional and the \ac{sps}-aided empirical estimation of the covariance for coherent sources. The results, depicted in Fig.~\ref{fig:eigenvalues}, demonstrate that \acs{ssn} yields a clear separation between the $M=3$ dominant eigenvalues of the signal subspace and those associated with noise. In contrast, conventional empirical estimation is only capable of identifying one eigenvalue associated with the signal subspace, whereas the \ac{sps} empirical estimation method can identify only two eigenvalues, with the remaining eigenvalues attributed to the noise subspace. The capability of \acs{ssn} to distinguish between the two subspaces significantly enhances resolving coherent sources compared to model-based subspace methods.

%
%

\subsubsection{Spectrum Presentation}
\label{sssec:Spectrum presentation}
\begin{figure*}
\begin{center}
\begin{subfigure}[pt]{0.32\linewidth}
\includegraphics[width=\linewidth]{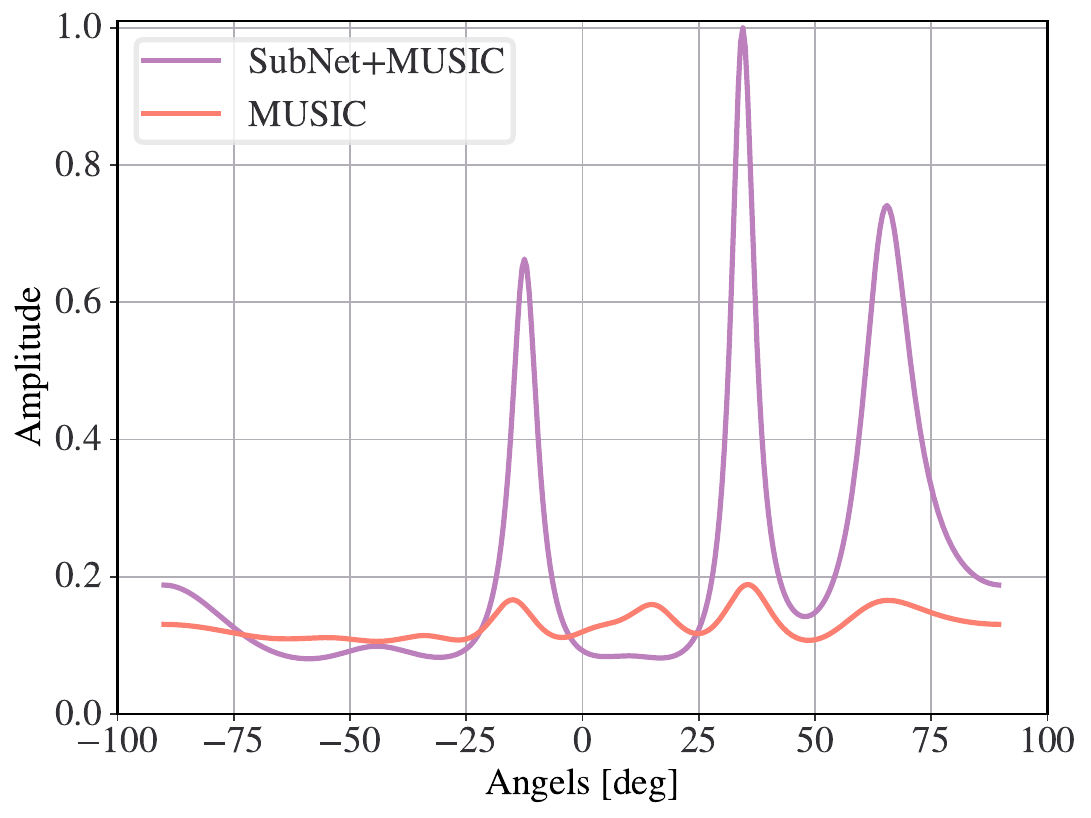}
\caption{\ac{music} $\&$ \acs{ssn}+\ac{music}}
\label{fig:music_NB_coherent_M_3}
\end{subfigure}
\begin{subfigure}[pt]{0.32\linewidth}
\includegraphics[width=\linewidth]
{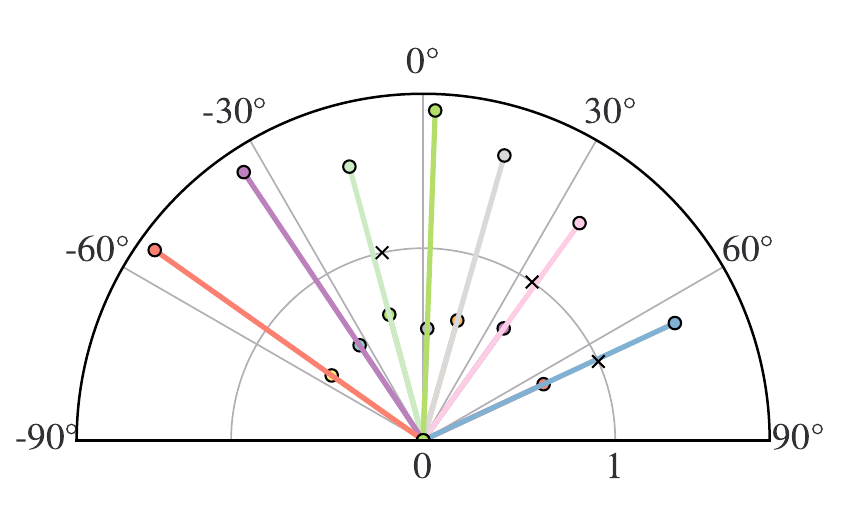}
\caption{\ac{rm}}
\label{fig:rm_NB_coherent_M_3}
\end{subfigure}
\begin{subfigure}[pt]{0.32\linewidth}
\includegraphics[width=\linewidth]{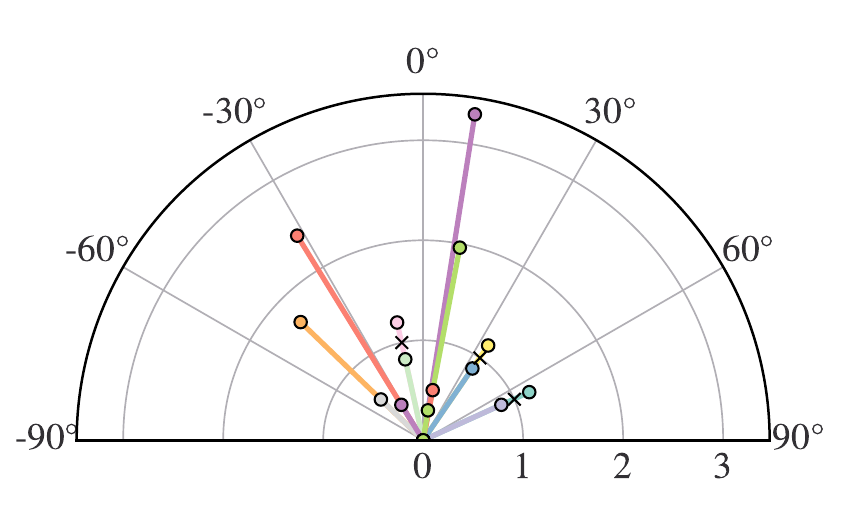}
\caption{\acs{ssn}+\ac{rm}}
\label{fig:ssn_rm_NB_coherent_M_3}
\end{subfigure}
\caption{Spectrum representation: $M=3$ narrowband coherent sources}%
\label{fig: spectrum_NB_coherent_M_3}
\vspace{0.2cm}
\begin{subfigure}[pt]{0.32\linewidth}
\includegraphics[width=\linewidth]{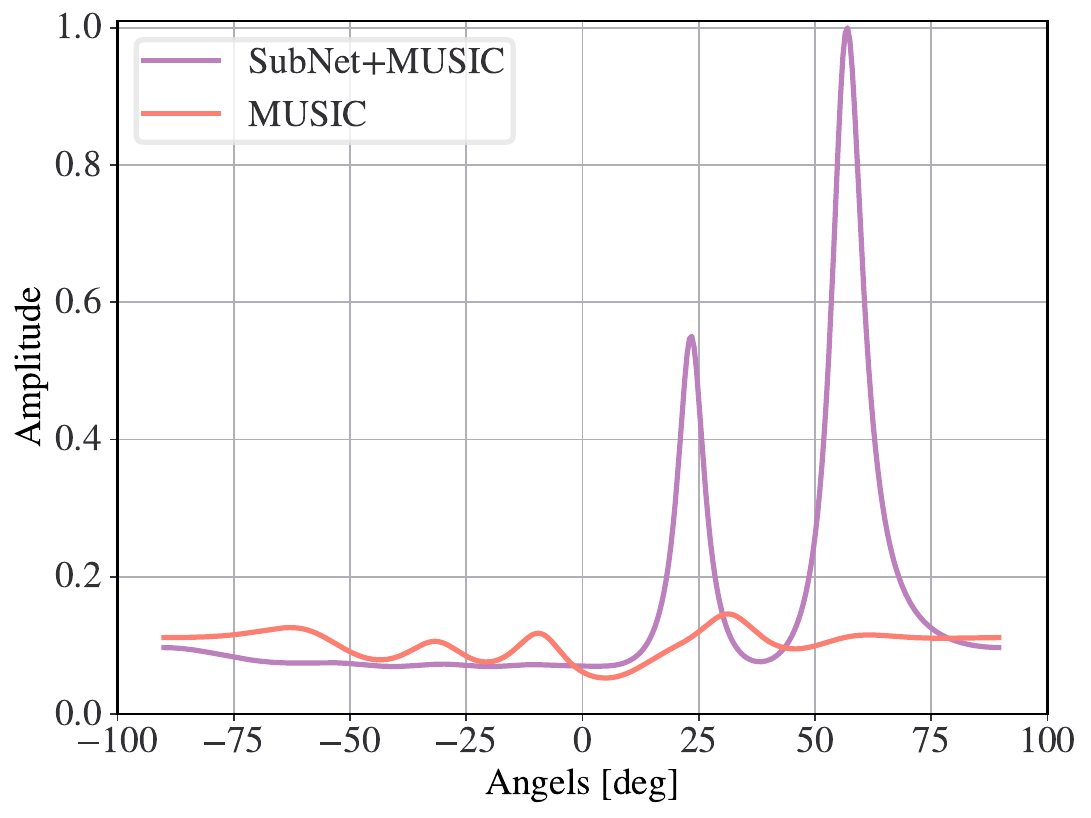}
\caption{\ac{music} $\&$ \acs{ssn}+\ac{music}}
\label{fig:music_NB_coherent_T_2}
\end{subfigure}
\begin{subfigure}[pt]{0.32\linewidth}
\includegraphics[width=\linewidth]{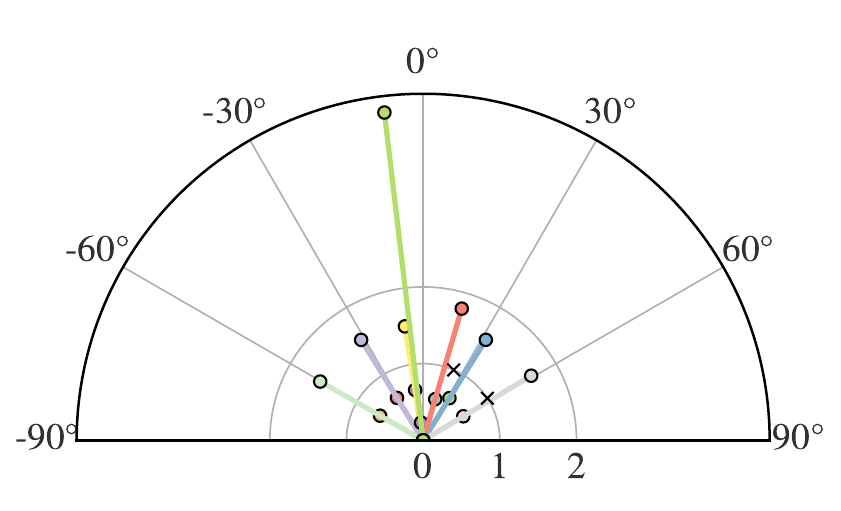}
\caption{\ac{rm}}
\label{fig:rm_NB_coherent_T_2}
\end{subfigure} 
\begin{subfigure}[pt]{0.32\linewidth}
\includegraphics[width=\linewidth]{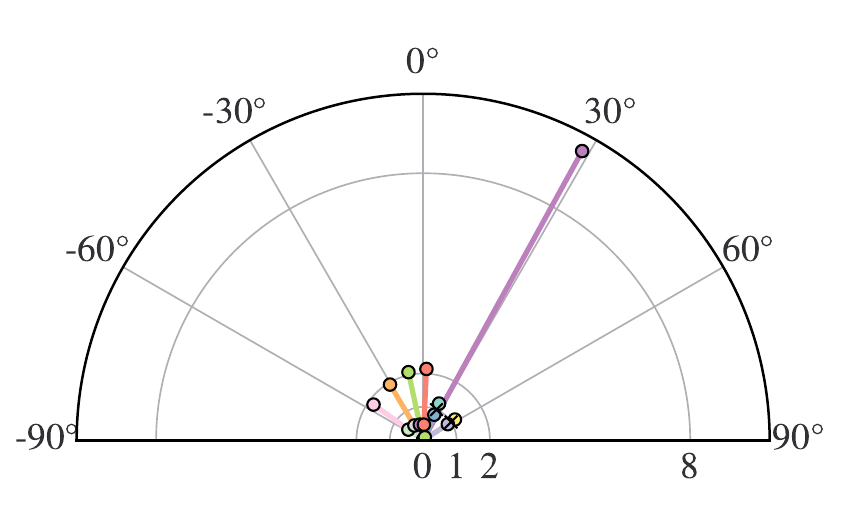}
\caption{\acs{ssn}+\ac{rm}}
\label{fig:ssn_rm_NB_coherent_T_2}
\end{subfigure} 
%
%
\caption{Spectrum representation: $M=2$ narrowband coherent sources with few snapshots $\&$ moderate \ac{snr}}%
\label{fig: spectrum_NB_coherent_T_2}
\vspace{0.2cm}
\begin{subfigure}[pt]{0.32\linewidth}
\includegraphics[width=\linewidth]{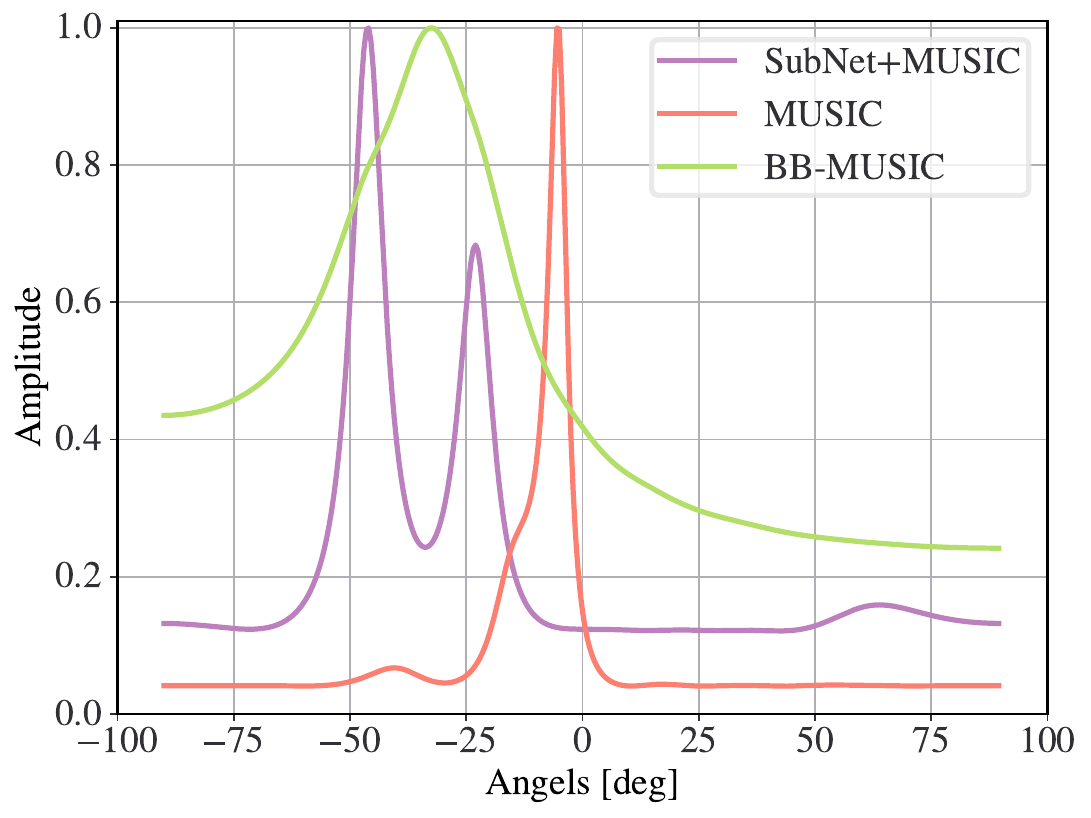}
\caption{\ac{music} $\&$ \acs{ssn}+\ac{music}}
\label{fig:music_BB_coherent}
\end{subfigure} 
\begin{subfigure}[pt]{0.32\linewidth}
\includegraphics[width=\linewidth]{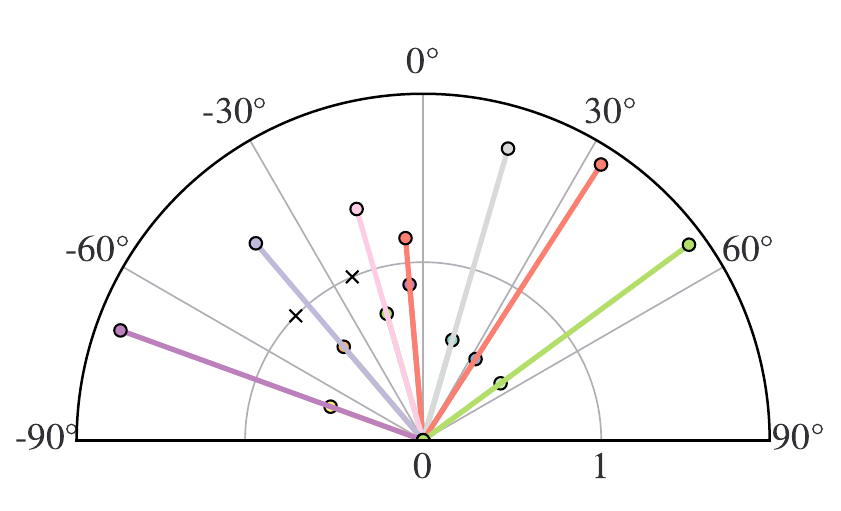}
\caption{Broadband $\&$ coherent, $T=50$}
\label{fig:rm_BB_coherent}
\end{subfigure} 
\begin{subfigure}[pt]{0.32\linewidth}
\includegraphics[width=\linewidth]{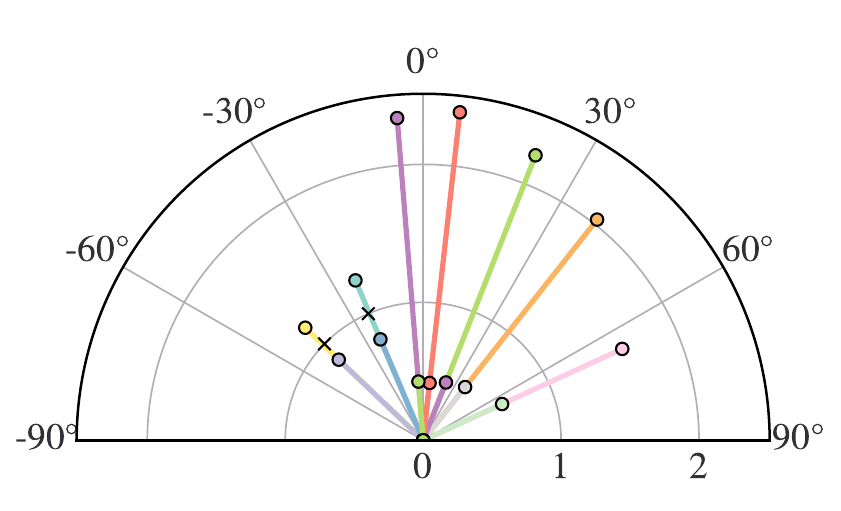}
\caption{Broadband $\&$ coherent, $T=50$}
\label{fig:ssn_rm_BB_coherent}
\end{subfigure} 
\caption{Spectrum representation: $M=3$ Wide-band coherent sources}%
\label{fig: spectrum_BB_coherent}
\vspace{-0.4cm}
\end{center}
\end{figure*}

When assumptions \ref{itm:narrowband} to \ref{itm:Snapshots} hold, subspace methods demonstrate informative spectral representations that are strongly correlated with the true \acp{doa}. Specifically, the \ac{music} spectrum displays distinct peaks at the \ac{doa} angles, while the \ac{rm} polar spectra exhibit roots and eigenvalues corresponding to the \ac{doa} locations on the unit circle.
We next investigate the spectral behavior of each subspace method and its \acs{ssn} augmentation, over the following scenarios in which \ref{itm:narrowband}-\ref{itm:Snapshots} are violated:
\begin{enumerate}
    \item  \ac{doa} estimation of narrowband coherent sources, i.e., \ref{itm:coherent} does not hold. The \acp{doa} are located at $\myVec{\theta} = [-12.34^{\circ}, 34.56^{\circ}, 65.78^{\circ}]$, and are observed via $T=100$ snapshots at \ac{snr} of $10$ dB. The resulting \ac{music} and \ac{rm} spectra are illustrated in Fig.~\ref{fig: spectrum_NB_coherent_M_3}.
    \item   Scenario where  \ref{itm:coherent} and \ref{itm:Snapshots} do not hold, using the same settings as in the study reported in Table~\ref{table:extreme scenario sim}, with sources at $\myVec{\theta} = [23.45^{\circ}, 56.78^{\circ}]$. The resulting \ac{music} and \ac{rm} spectra are illustrated in Fig.~\ref{fig: spectrum_NB_coherent_T_2}.
    \item   \ac{doa} estimation of broadband coherent sources, located at $\myVec{\theta} = [-45.67^{\circ}, -23.45^{\circ}]$, from $T=50$ snapshots.
The \acl{mb} spectrum here is no longer informative in representing the true \ac{doa}, as illustrated in Fig.~\ref{fig: spectrum_BB_coherent}.
\end{enumerate}

We observe in Figs.~\ref{fig: spectrum_NB_coherent_M_3}-\ref{fig: spectrum_BB_coherent}, that  \acs{ssn}  produces an interpretable and informative spectrum. The roots closest to the unit circle in the \acs{ssn}+\ac{rm} spectrum (Figs.~\ref{fig: spectrum_NB_coherent_M_3}\subref{fig:ssn_rm_NB_coherent_M_3}-\ref{fig: spectrum_BB_coherent}\subref{fig:ssn_rm_BB_coherent}) and the distinct peaks in the \acs{ssn}+\ac{music} spectrum (Figs.~\ref{fig: spectrum_NB_coherent_M_3}\subref{fig:music_NB_coherent_M_3}-\ref{fig: spectrum_BB_coherent}\subref{fig:music_BB_coherent}) can be easily linked to the actual \acp{doa}. On the other hand, classical \ac{music} and \ac{rm} algorithms are unable to produce meaningful spectrum representations. While \ac{music} spectrum for narrowband coherent sources (Fig.~\ref{fig: spectrum_NB_coherent_M_3}\subref{fig:music_NB_coherent_M_3}) may reveal minor peaks near the source angles, it horribly fails to provide any meaningful information for the scenarios depicted Figs.~\ref{fig: spectrum_NB_coherent_T_2}\subref{fig:music_NB_coherent_T_2}-\ref{fig: spectrum_BB_coherent}\subref{fig:music_BB_coherent}. 
Furthermore,  \acs{ssn} enables a more distinguishable spectrum. As shown in Figs.~\ref{fig: spectrum_NB_coherent_M_3}\subref{fig:ssn_rm_NB_coherent_M_3}-\ref{fig: spectrum_BB_coherent}\subref{fig:ssn_rm_BB_coherent}, the non-\ac{doa} roots are pushed further away from the unit circle, facilitating their distinction from true \ac{doa} roots. The resulting \ac{rm} spectrum becomes more interpretable , especially in challenging scenarios. This  is clearly observed in the scenario with a small number of snapshots in Fig.~\ref{fig: spectrum_NB_coherent_T_2}\subref{fig:ssn_rm_NB_coherent_T_2}, where a non-\ac{doa} root which located close to the true \ac{doa}, at $\myScal{\theta}=28.4^{\circ}$, is pushed from the unit circle ($\abs{\rm root} > 8$), preventing misclassification.


\subsubsection{Beamforming Pattern}
\label{sssec:Beamforming pattern}
We conclude by demonstrating that the surrogate covariance produced by \acs{ssn} is useful not only for subpsace methods, but also for other forms of covariance-based array  processing algorithms. To exemplify this, we consider \ac{mvdr}~\cite{capon1969high}, which is a popular beamforming method that  enhances the  signal power in the desired direction while suppressing  noise and interference. This results in a beampattern that provides valuable information about the directions of the sources of interest. Although  \ac{mvdr}  does not rely on subspace decomposition for its beamforming operation, it employs the covariance matrix to shape its pattern, and hence it is subject to the same empirical estimation limitations as classical subspace methods, and its effectiveness is still limited by the accuracy of the covariance matrix estimate.

\begin{figure*}
\begin{center}
\begin{subfigure}[pt]{0.32\linewidth}
\includegraphics[width=\linewidth]{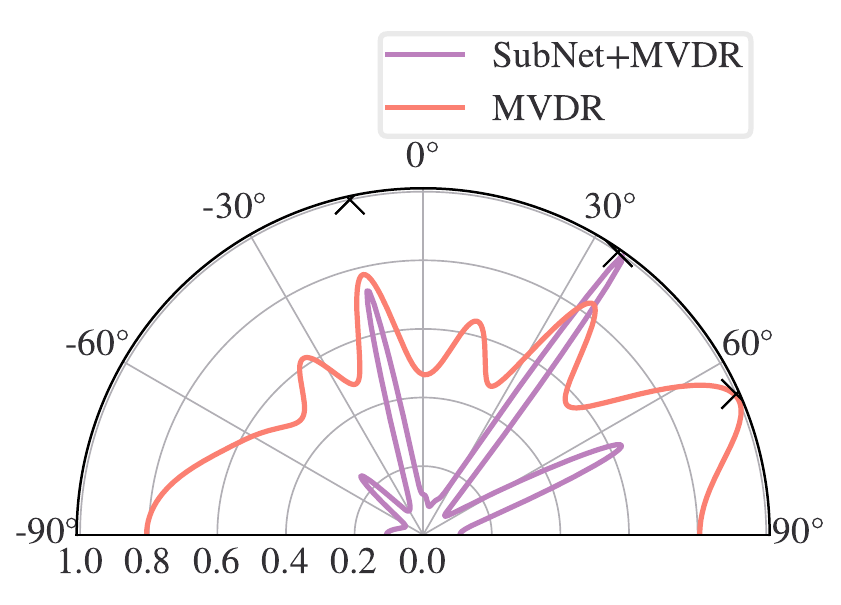}
\caption{narrowband $\&$ coherent, $M=3$.}
\label{fig:mvdr_M_3_coherent}
\end{subfigure}
\begin{subfigure}[pt]{0.32\linewidth}
\includegraphics[width=\linewidth]{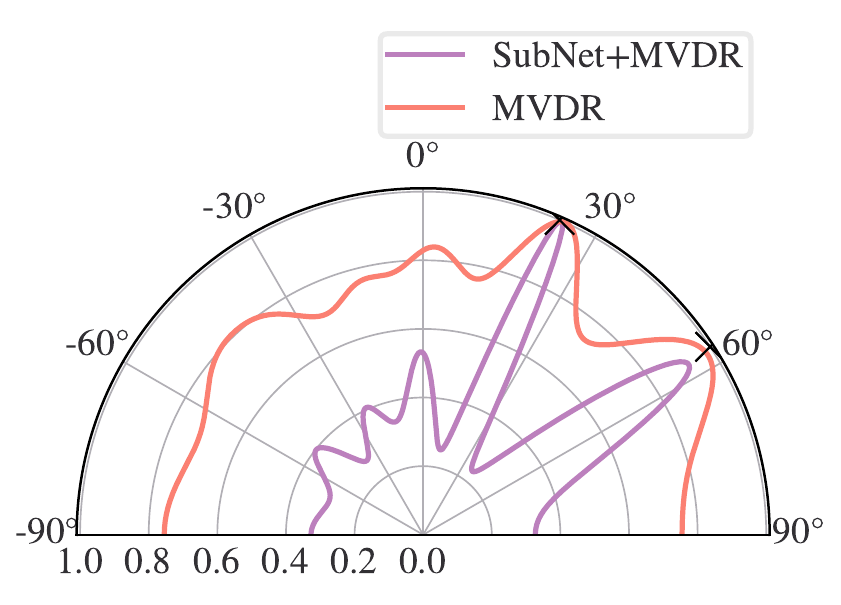}
\caption{narrowband $\&$ coherent, low \ac{snr}.}
\label{fig:mvdr_coherent_low_snr}
\end{subfigure} 
\begin{subfigure}[pt]{0.32\linewidth}
\includegraphics[width=\linewidth]{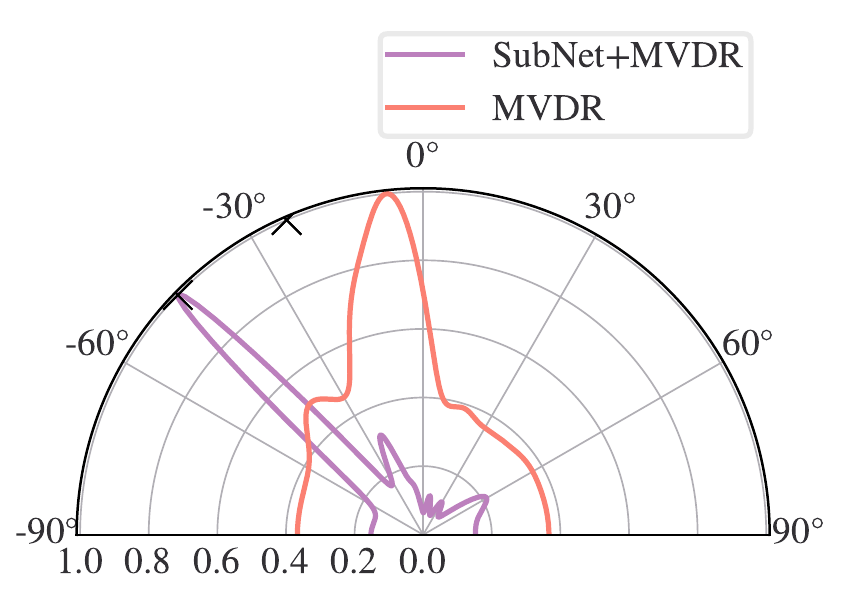}
\caption{Broadband $\&$ coherent}
\label{fig:mvdr_BB_coherent}
\end{subfigure} 
\caption{\acs{ssn} $\&$ \ac{mvdr} beampattern comparison}
\label{fig:mvdr_spectrum}
\end{center}
\vspace{-0.2cm}
\end{figure*}

To show that \acs{ssn} facilitates the operation of \ac{mvdr}, we visualize the beampattern achieved by \ac{mvdr} with and without \acs{ssn} augmentation in Fig.~\ref{fig:mvdr_spectrum}, under the same settings used in Subsection~\ref{sssec:Spectrum presentation}. 
The beampatterns depicted in Fig.~\ref{fig:mvdr_spectrum} clearly demonstrate the superiority of \acs{ssn} in producing surrogate covariance matrices that are useful not only for \ac{doa} estimation subspace methods, but also to obtain high resolution effective beampatterns. 
Specifically, Figs.~\ref{fig:mvdr_spectrum}\subref{fig:mvdr_M_3_coherent}-\ref{fig:mvdr_spectrum}\subref{fig:mvdr_coherent_low_snr} demonstrate that \acs{ssn} produces narrow lobes in the direction of the sources, enhancing the discrimination between sources, and mitigating interference by reducing the power  in non-\ac{doa} directions. When using the conventional empirical covariance, \ac{mvdr} exhibits limitations in achieving these objectives in scenarios with multiple coherent sources and low \ac{snr}. Additionally, like subspace methods, \ac{mvdr} is inherently limited in its ability to  handle broadband sources. This shortcoming leads to a beampattern that deviates significantly from the true \acp{doa}, as shown in Fig.~\ref{fig:mvdr_spectrum}\subref{fig:mvdr_M_3_coherent}. Nonetheless, by integrating the \acs{ssn} augmentation, the received signal power in the desired directions is amplified, which enhances the effectiveness of beamforming for signals with both broadband and coherent characteristics.
%
%
\vspace{-0.2cm}
\section{Conclusions}
\label{ssec:conclusions}
\vspace{-0.1cm}
{We proposed \acs{ssn}, which enables subspace-based \ac{doa} estimation in challenging settings where a reliable division of the array input into signal and noise subspaces is typically infeasible. \acs{ssn} is designed by augmenting subspace methods with deep learning tools, and it is particularly trained to produce a surrogate covariance that is universally useful for subspace-based \ac{doa} estimation by identifying \ac{rm} as a suitable differentiable method and converting it into a trainable architecture. \acs{ssn} leverages data to learn to divide the observations into signal and noise subspaces. Our results demonstrate its ability to successfully cope with coherent sources, broadband signals, few snapshots, low \ac{snr}, and array miscalibrations, while preserving the interpretability of model-based subspace methods, and facilitating the operation of other covariance-based algorithms such as \ac{mvdr}.}
\bibliographystyle{IEEEtran}
\bibliography{IEEEabrv,DoA}

\end{document}